\documentclass[a4paper,11pt,nofootinbib,floatfix]{revtex4-2}

\usepackage{amssymb,bm,epsf}
\usepackage{setspace}
\usepackage{amsmath}
\usepackage{graphicx}
\usepackage{dcolumn}
\usepackage{color}
\usepackage{accents}
\usepackage{multirow}
\usepackage{tikz}
\usepackage{comment}
\usepackage{subfigure}
\usepackage{soul}
\usepackage[rightcaption]{sidecap}
\sidecaptionvpos{figure}{t}
\usepackage{enumerate}

\usepackage[toc,page]{appendix}
\usepackage[normalem]{ulem}

\usepackage[hyperfootnotes=false]{hyperref}
\hypersetup{
    colorlinks=true,
    linkcolor=red,
    filecolor=magenta,      
    citecolor=blue
}

\usepackage{titlesec}

\titleformat{\section}
    {\normalfont\bfseries\normalsize\centering}
    {\thesection}{1em}{\MakeUppercase}
\titlespacing{\section}
    {0pt}   
    {2.0ex plus .5ex minus .2ex}  
    {0.8ex plus .1ex}             

\titleformat{\subsection}
    {\normalfont\bfseries\centering}{\makebox[1cm][l]{\thesubsection}}{0pt}{}
\titlespacing{\subsection}
    {0pt}
    {1.5ex plus .2ex minus .2ex}
    {0.8ex plus .1ex}

\newcolumntype{P}[1]{>{\centering\arraybackslash}p{#1}}
\newcolumntype{M}[1]{>{\centering\arraybackslash}m{#1}}

\makeatletter
\@addtoreset{equation}{section}

\makeatother

\begin{document}
\begin{titlepage}
\title{The Graviton Propagator in Asymptotically Safe Gravity with Non-Local Form Factors}
\author{Emiliano Maria Glaviano}
\email{emiliano.glaviano@inaf.it}
\affiliation{
INAF Osservatorio Astrofisico di Catania, Via S.Sofia 78, 95123 Catania ITALY}
\affiliation{INFN, Sezione di Catania, Via S.\ Sofia 64, 95123 Catania, Italy}
\affiliation{Università di Catania, Dipartimento di Fisica e Astronomia, Via S. Sofia 64, 95123, Catania, Italy}


\begin{abstract}
\noindent We analyze the flow of the background graviton propagator driven by the running of asymptotically safe form factors in four-dimensional quantum gravity at quadratic order in the curvature expansion. We construct the propagator in the  $k=0$ limit and investigate its momentum dependence. The non-local form factors are then analytically continued to Minkowski spacetime, from which the corresponding Minkowskian propagator is obtained. We find a single pole at $q^2=0$ with positive residue and no additional ghost poles at this level of approximation. The corresponding Newtonian potential is found to be regular at $r=0$. 
\end{abstract}

\maketitle

\end{titlepage}
\newpage
\setcounter{page}{2}

\section{Introduction}\label{introduction}
\noindent Propagators play a central role in quantum field theory. They constitute the basic building blocks of the effective action, encode unitarity properties, and in the non-relativistic limit are directly related to the static potential \cite{Peskin:1995ev,Weinberg:1995mt,Weinberg:1996kr}. In gravity, however, its interpretation is considerably more subtle. General relativity is perturbatively non-renormalizable and quantum corrections generate higher-derivative and non-local contributions to the effective action \cite{Donoghue:1993eb,Donoghue:1994dn,Holstein:2004dn,Akhundov:1996jd,Codello:2015mba}. Understanding how the non-local interactions affect the graviton propagator within a consistent quantum theory of gravity is an open problem.

Several approaches have been proposed to build a consistent theory of quantum gravity, including string theory \cite{Sen:2024nfd}, supergravity \cite{Taylor:1983su}, and the asymptotic safety scenario \cite{Basile:2024oms,Eichhorn:2017egq,Bonanno:2020bil,Niedermaier:2006ns,Reuter:2012id,Dupuis:2020fhh,article,book,Saueressig:2023irs,Reuter_Saueressig_2019}. Non-local modifications of the Einstein--Hilbert action arise naturally in string theory \cite{FREUND1987186,Witten:1985cc,Biswas:2004qu} and in renormalization group analyses \cite{Knorr:2019atm,Knorr:2021iwv,Knorr:2022dsx}. In a curvature expansion, they are encoded in non-local form factors \cite{Barvinsky:1990up,Barvinsky1987BeyondTS,Barvinsky:1990uq,Barvinsky:1993en,Avramidi:2000bm}, which generically lead to theories with infinitely many derivatives.

Gravitational theories with infinitely many derivatives exhibit improved ultraviolet behavior compared to local higher-derivative models \cite{Biswas:2011ar,Frolov:2015usa,Biswas:2013cha,Biswas:2013kla,Modesto:2011kw,Tomboulis:1997gg,BasiBeneito:2022wux,Modesto:2017sdr,Biswas:2010zk,Biswas:2005qr}. In static weak-field configurations, this manifests in a regular Newtonian potential at $r=0$ \cite{Biswas:2011ar,Biswas:2005qr,Modesto:2011kw,Frolov:2015usa,Giacchini:2016xns,Burzilla:2020bkx,Giacchini:2018wlf}, while in curved spacetimes it is associated with regular black-hole solutions \cite{Biswas:2011ar,Frolov:2015usa,Modesto:2011kw,Buoninfante:2018xiw,Buoninfante:2018rlq,Burzilla:2020utr,Buoninfante:2022ild,Knorr:2021iwv,Burzilla:2020utr,Burzilla:2026uzv,Buoninfante:2018stt}.  Requiring that the non-local form factors are entire functions could also alleviate the ghost problem \cite{Biswas:2011ar,Frolov:2015usa,Biswas:2013cha,Biswas:2013kla,Modesto:2011kw,Tomboulis:1997gg,BasiBeneito:2022wux,Modesto:2017sdr,Biswas:2010zk,Biswas:2005qr,Buoninfante:2020ctr,Modesto:2012ys}. 

In the asymptotic safety scenario  non-local structures arise dynamically from the RG flow of the effective action. The study of non-local form factors within this framework has seen significant developments in recent years \cite{Knorr:2019atm,Bosma:2019aiu,Knorr:2022dsx,Knorr:2021niv,Knorr:2022kqp,Knorr:2018kog}. Within curvature-squared truncations, form factors avoid additional poles in the graviton propagator \cite{Knorr:2021niv}, while in conformally reduced gravity they regularize the Newtonian potential, removing the $1/r$ singularity \cite{Bosma:2019aiu,Knorr:2021iwv}. 

Momentum-dependent correlation functions in asymptotically safe gravity have been extensively studied \cite{article,Christiansen:2012rx,Christiansen:2014raa,Christiansen:2015rva,Denz:2016qks,Bonanno:2021squ} using a vertex expansion for the effective action.
The dressed anomalous dimensions decay to zero at large momentum implying momentum locality and absence of quadratic divergences in the flow of $\Gamma_k^{(2)}(q^2)$. It has also been observed that graviton $n$-point couplings exhibit an approximate universality \cite{Eichhorn:2018ydy,Eichhorn:2018akn,Eichhorn:2018nda}. The momentum dependence of the graviton propagator can be also encoded in an effective Newton coupling \cite{Bonanno:2021squ,Pawlowski:2023dda,Denz:2016qks}, whose $q^2$ dependence qualitatively resembles the RG running of $G_k$ under the identification $k^2 \leftrightarrow q^2$. 

Physical consequences of the graviton propagator have been investigated through scattering amplitudes \cite{Chiesa:2026tlz, Knorr:2026vax, Knorr:2022lzn, Pastor-Gutierrez:2024sbt,Draper:2020knh} and the spectral representation \cite{Pawlowski:2025etp,Bonanno:2021squ,Fehre:2021eob,Pastor-Gutierrez:2024sbt,Kher:2025rve,article} inside the Einstein-Hilbert truncation. The background spectral density becomes negative at high energies, while the fluctuation spectral density remains positive at all scales. 

A remaining challenge concerns the analytic continuation of Euclidean results to Minkowski space. Several approaches have been proposed in the literature \cite{Eichhorn:2019ybe,Nagy_2019,Baldazzi:2018mtl,Knorr:2018fdu,Wetterich:2017ixo,Houthoff:2017oam,Biemans:2016rvp,Demmel:2015zfa,Rechenberger:2012dt,Manrique:2011jc} but a fully general solution is still under development.

More recently, progress on non-local form factors has been achieved using the proper-time formalism \cite{Bonanno:2019ukb,Glaviano:2026lew,Glaviano:2024hie,Bonanno:2026ljx}. In \cite{Glaviano:2026lew}, the flow of asymptotically safe background gravitational form factors at second order in the curvature expansion was computed and the corresponding solutions at $k=0$ were obtained. These exhibit the characteristic infrared and ultraviolet scaling behavior expected in asymptotically safe gravity \cite{Knorr:2019atm,Knorr:2022dsx}.

In this work, building on \cite{Glaviano:2026lew}, we study the impact of asymptotically safe form factors on the graviton propagator and investigate whether known properties of the graviton two-point function can be derived directly from their RG flow. We further analytically continue the Euclidean results to Minkowski spacetime and compute the corrections to the Newtonian potential. While the propagator retains the standard massless pole structure, non-local effects significantly modify its infrared and ultraviolet behavior, leading to a regular short-distance potential and deviations from the $1/r$ law at large distances.

Although exploratory, these results provide insight into the physical implications of asymptotically safe form factors. The present analysis, based on a background approximation, neglects backreaction effects and fully dynamical spacetime fluctuations and should therefore be regarded as a first step toward a more complete description of the resulting non-local dynamics.

The paper is organized as follows. In Sec.~\ref{secform}, we summarize the main results of \cite{Glaviano:2026lew} relevant for the present work. In Sec.~\ref{seceucprop} we analyze the corrections to the Euclidean graviton propagator induced by the form factors at $k=0$. In Sec.~\ref{secEFFgamma2} we extend the discussion to include the running of the form factors and analyze the resulting two-point function. Sec.~\ref{secAnalcont} is devoted to the analytic continuation of the asymptotically safe form factors to Minkowski spacetime, while Sec.~\ref{secminkprop} addresses the properties of the corresponding Minkowski propagator. In Sec.~\ref{secNewtpot} we compute the corrections to the Newtonian potential. Finally, Sec.~\ref{secconc} contains our conclusions. Three appendices complement the main text: Appendix~\ref{appderbeta} outlines the derivation of the Hessian, Appendix~\ref{appNewpot} discusses the technique to derive the asymptotics of the Newtonian potential without performing the full Fourier transform of the graviton propagator, and Appendix~\ref{App:asymFF} collects technical results from \cite{Glaviano:2026lew} used throughout the paper.

\section{Properties of Asymptotically Safe Form Factors}\label{secform}
\noindent In this section we briefly review the results of \cite{Glaviano:2026lew} relevant for the present analysis.

The starting point of \cite{Glaviano:2026lew} is the Euclidean effective action
\begin{equation}
\label{TE}
\Gamma_k[g] = \int d^d x \sqrt{g} \left[\frac{1}{16\pi G_k}\left(-R + 2\bar{\lambda}_k\right) + R f_k^{(\mathrm{R})}(-\Box) R + R_{\mu\nu} f_k^{(\mathrm{Ricc})}(-\Box) R^{\mu\nu}\right]
\end{equation}
where $-\Box$ is the Laplace--Beltrami operator, while $f_k^{(\mathrm{R})}(-\Box)$ and $f_k^{(\mathrm{Ricc})}(-\Box)$ are running form factors. The flow of the background form factors was derived within the proper-time formalism using two different prescriptions for the regulator, the mixed and the B-scheme; technical details are given in Appendix~\ref{App:asymFF}. At $k=0$, both schemes lead to the same functional structure of the form factors, differing only in their numerical values. The mixed scheme allows for analytic control of the asymptotic regimes, whereas the full B-scheme requires numerical evaluation. In this work, we set $\bar\lambda_k=0$ and define the Planck mass at $k=0$ as $m_p=1/G_{k=0}$.

The running of the asymptotically safe form factor is given by
 \begin{equation}
 \label{ASflow}
\bar F_{k}^{(i,\mathrm{AS})}(x;x_{\mathrm{BC}})=F_{k}^{(i,\mathrm{AS})}(x;x_{\mathrm{BC}})+h(x_{\mathrm{BC}}),
\end{equation}
where $i=\mathrm{R},\mathrm{Ricc}$ and $x=q^2/k^2$. Here $F_{k}^{(i,\mathrm{AS})}$ denotes the reference solution, while $h(x_{\mathrm{BC}})$ encodes the boundary conditions of the fixed-point equation, with $x_{\mathrm{BC}}$ an arbitrary dimensionless scale. The shift $h(x_{\mathrm{BC}})$ is fixed by specifying the asymptotic value of the form factor at infinite momentum, $x_{\mathrm{BC}}\to\infty$ \cite{Knorr:2021niv}. The explicit expression for $F_{k}^{(i,\mathrm{AS})}$ is provided in Appendix~\ref{App:asymFF}.

\begin{figure}[t]
   \centering
     \subfigure[]{
         \centering
         \includegraphics[width=0.45\textwidth]{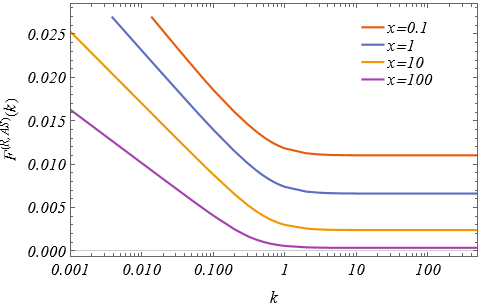}
     }
    \hspace{0.8em}
     \subfigure[]{
         \centering
         \includegraphics[width=0.45\textwidth]{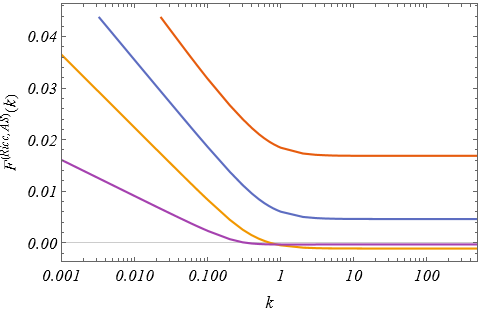}
     }
     \caption{Flow of the asymptotically safe solution in Eq.~(\ref{ASflow}) for different values of $x$. Panels (a) and (b) show $F_k^{(\mathrm{R})}(x)$ and $F_k^{(\mathrm{Ricc})}(x)$, respectively. Technical details are given in App. \ref{App:asymFF}.}
     \label{plotflowad}

\vspace{0.5cm}

     \centering
     \subfigure[]{
         \centering
         \includegraphics[width=0.45\textwidth]{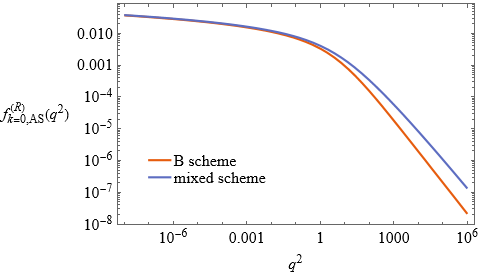}
     }
    \hspace{0.8em}
     \subfigure[]{
         \centering
         \includegraphics[width=0.45\textwidth]{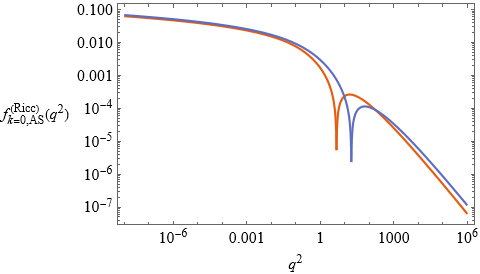}
     }
     \caption{Reference AS form factors at $k=0$ obtained from Fig.~\ref{plotflowad}. Panel (a) and (b) show $f_\mathrm{AS}^{(\mathrm{R})}(q^2)$ and $f_\mathrm{AS}^{(\mathrm{Ricc})}(q^2)$ respectively. The qualitative behavior is scheme independent.}
    \label{plotFFASeps1}
\end{figure}

The resulting flow interpolates between a Gaussian regime for $k\ll m_p$, reproducing the EFT behaviour, and a non-Gaussian fixed-point regime for $k\gg m_p$. This crossover is illustrated in Fig.~\ref{plotflowad} for representative values of $x$. In the vicinity of the two regimes, the solutions admit expansions of the form
\begin{equation}\begin{split}
\label{serieGFPNGFP}
&F_k^{\mathrm{GFP}}\left(x\right)=F_\ast^{\mathrm{GFP}}\left(x\right)+\sum_{n=1}^{+\infty}{F_n^{\mathrm{GFP}}\left(x\right)\left(\frac{k}{m_p}\right)^{2n}}+F_B\left(x,\frac{k^2}{m_p^2}\right)\\
&F_k^{\mathrm{NGFP}}\left(x;x_{\mathrm{BC}}\right)=F_\ast^{\mathrm{NGFP}}\left(x;x_{\mathrm{BC}}\right)+\sum_{n=1}^{+\infty}{F_n^{\mathrm{NGFP}}\left(x\right)\left(\frac{m_p}{k}\right)^{2n}}
\end{split}\end{equation}
The leading and the higher-order terms correspond to the fixed-point solutions and perturbations around them respectively. $F_B$ encodes the non-stationary component of the infrared RG flow. Representative expressions for the perturbations and details on $F_B$ are given in Appendix~\ref{App:asymFF}.

The physical form factors are obtained by taking the $k\to0$ limit of the running solutions. In dimensional variables, they read
\begin{equation}
\label{FFk=0}
\bar f_{k=0}^{(i,\mathrm{AS})}(q^2)\equiv \lim_{\substack{k\to0\\ x_{\mathrm{BC}}\to\infty}}\bar F_{k}^{(i,\mathrm{AS})}\left(\frac{q^2}{k^2};x_{\mathrm{BC}}\right)=f_{k=0}^{(i,\mathrm{AS})}(q^2)+c_i,
\end{equation}
where we set $h(x_{\mathrm{BC}}\to\infty)=c_i$. The reference numerical solutions are shown in Fig.~\ref{plotFFASeps1}. 

To characterize the form factors, it is convenient to introduce the dimensionless variable
\begin{equation}
\label{varadim}
w=\frac{q^2}{m_p^2},
\end{equation}
such that the infrared (IR) and ultraviolet (UV) regimes correspond to $w\ll1$ and $w\gg1$, respectively. In the IR, the reference asymptotically safe form factors admit the expansion
\begin{equation}
\label{formASIR}
f_{k=0}^{\left(i,\mathrm{AS}\right)}(w\rightarrow0)=\sum_{n=0}^{+\infty}{\left[a_n^{\left(i\right)}+b_n^{\left(i\right)}\ln{\left(\frac{23w}{12\pi}\right)}\right]w^n}
\end{equation}
where $i=R,Ricc$. In the mixed scheme, the leading coefficients read
\begin{equation}
\label{coeffIRk0expIR}
\begin{aligned}
&a_0^{(\mathrm{R})}= \frac{287}{11520\pi^2},&\qquad
&b_0^{(\mathrm{R})}= -\frac{7}{384\pi^2},&\qquad
&a_1^{(\mathrm{R})}= \frac{31257}{2508800\pi^3},&\qquad
&b_1^{(\mathrm{R})}= -\frac{299}{35840\pi^3},\\[1mm]
&a_0^{(\mathrm{Ricc})}= -\frac{103}{5760\pi^2},&\qquad
&b_0^{(\mathrm{Ricc})}= -\frac{7}{192\pi^2},&\qquad
&a_1^{(\mathrm{Ricc})}= \frac{381547}{11289600\pi^3},&\qquad
&b_1^{(\mathrm{Ricc})}= \frac{3013}{53760\pi^3}.
\end{aligned}
\end{equation}
The leading behavior is logarithmic, as inherited from the EFT regime, while subleading corrections are organized as powers of $w$ accompanied by logarithmic terms. In contrast, the UV behavior of the reference solutions is described by
\begin{equation}
\label{formASUV}
f_{k=0}^{\left(i,\mathrm{AS}\right)}(w\rightarrow\infty)=\sum_{n=1}^{+\infty}\frac{a_n^{\left(i\right)}+b_n^{\left(i\right)}\ln{\left(\frac{23w}{12\pi}\right)}}{w^n}
\end{equation}
The first terms of these expansions are given by
\begin{equation}\begin{aligned}
&a_1^{\left(\mathrm{R}\right)}=-\frac{43}{1104\pi},&\qquad &b_1^{\left(\mathrm{R}\right)}=\frac{25}{736\pi},&\qquad
&a_2^{\left(\mathrm{R}\right)}=\frac{9}{2116},&\qquad
&b_2^{\left(\mathrm{R}\right)}=\frac{81}{2116},&\qquad\\
&a_1^{\left(\mathrm{Ricc}\right)}=\frac{137}{1104\pi},&\qquad &b_1^{\left(\mathrm{Ricc}\right)}=-\frac{13}{368\pi},&\qquad
&a_2^{\left(\mathrm{Ricc}\right)}=\frac{9}{529},&\qquad
&b_2^{\left(\mathrm{Ricc}\right)}=-\frac{117}{1058}
\end{aligned}\end{equation}
As a consequence of the asymptotically safe fixed-point structure, the form factors decay at large momentum, following an expansion in inverse powers of $w$ with logarithmic corrections.

The flow of the form factors will be used in Sec.~\ref{secEFFgamma2} to describe the running of the graviton two-point function, whereas the $k=0$ solutions will be used to describe the Euclidean propagator and its analytical continuation. The asymptotic expressions will provide analytical control in the infrared and ultraviolet regimes.

\section{Euclidean propagators}\label{seceucprop}
\noindent The propagator is obtained by inverting the Hessian of the theory in Eq.~(\ref{TE}), whose derivation is summarized in Appendix~\ref{appderbeta}. In a generic curved background, this inversion is highly non-trivial. We therefore restrict the analysis to a flat background, where the Hessian simplifies considerably.

In Lorentz gauge, the Hessian can be written as 
\begin{equation}
H^{cd}_{ab}(-\Box)=\left\{\delta_a^c\delta_b^d\,a_k(-\Box)+\delta^{cd}\left[b_k(-\Box)\,\partial_a\partial_b-\delta_{ab}\,c_k(-\Box)\right]\Box\right\}\Box
\end{equation}
where
\begin{equation}\begin{split}
&a_k(-\Box)=\frac{1}{16\pi G_k}-f_k^{(\mathrm{Ricc})}(-\Box)\,\Box,\quad\quad 
b_k(-\Box)=2f_k^{(\mathrm{R})}(-\Box)+f_k^{(\mathrm{Ricc})}(-\Box)\,,\\
&c_k(-\Box)=\frac{2}{2-d}f_k^{(\mathrm{R})}(-\Box)+\frac{1}{2-d}f_k^{(\mathrm{Ricc})}(-\Box)\,.
\end{split}\end{equation}
Decomposing into spin components, only the spin-2 and spin-0 sectors contribute, so that
\begin{equation}
H_{\mu\nu\alpha\beta}(q^2)
= a(q^2)\,P_{2,\mu\nu\alpha\beta}
+\left[a(q^2)q^2+(d-1)c(q^2)q^4\right]P_{S,\mu\nu\alpha\beta}\,.
\end{equation}
where $P_{2,\mu\nu\alpha\beta}$ and $P_{S,\mu\nu\alpha\beta}$ denote the Barnes--Rivers projectors \cite{Barnes:1965ylk,Rivers:1964nfl,VanNieuwenhuizen:1973fi} onto the spin-2 and spin-0 components, respectively.
The operator $H_{\mu\nu\alpha\beta}^{-1}$ is diagonal in the spin decomposition,
\begin{equation}\begin{split}
\label{prop}
H_{\mu\nu\alpha\beta}^{-1}(q^2)\equiv G_{\mu\nu\alpha\beta}(q^2)=G_{TT}(q^2)\,P_{2,\mu\nu\alpha\beta}+ G_{\mathrm{SS}}(q^2)\,P_{S,\mu\nu\alpha\beta},
\end{split}\end{equation}
with
\begin{equation}
\label{propTTSS}
G_{TT}(q^2)=\frac{1}{\frac{q^2}{16\pi G_k}+q^4 f_k^{(\mathrm{Ricc})}(q^2)}, \quad\quad
G_{\mathrm{SS}}(q^2)=\frac{1}{\frac{q^2}{16\pi G_k}+
\left(\frac{2(d-1)}{2-d}f_k^{(\mathrm{R})}(q^2)+\frac{1}{2-d}f_k^{(\mathrm{Ricc})}(q^2)\right)q^4}.
\end{equation}
The physical propagating contribution is encoded in the spin-2 sector, while the scalar sector is gauge-dependent and does not correspond to a propagating degree of freedom.

\subsection{Properties of the Euclidean propagator}
Using Eqs.~(\ref{formASIR}), the IR behavior of the propagator can be extracted. This is given by: 
\begin{equation}\begin{split}
\label{propcrcriccsmallp}
G_X\left(q^2,w\rightarrow0\right)=\frac{16\pi}{m_p^2q^2}\left[1+\sum_{n=1}^{+\infty}{\sum_{j=0}^{n}{a_{nj}^{\left(X\right)}\ln^j{\left(\frac{23w}{12\pi}\right)}}w^{n}}\right]
\end{split}\end{equation}
where $X=\mathrm{TT}$, $\mathrm{SS}$ and $a_{nj}^{\left(X\right)}$ are the series coefficients. The first coefficients are
\begin{equation}
\label{IRcoeffeucprop}
\begin{aligned}
a_{10}^{\mathrm{TT}}&=\frac{103}{360\pi}, \quad&
a_{11}^{\mathrm{TT}}&=\frac{7}{12\pi}, \quad&
a_{20}^{\mathrm{TT}}&=-\frac{1457041}{3175200\pi^2}, \quad&
a_{21}^{\mathrm{TT}}&=\frac{37211}{30240\pi^2}, \quad&
a_{22}^{\mathrm{TT}}&=\frac{49}{144\pi^2},
\\
a_{10}^{\mathrm{SS}}&=\frac{379}{360\pi}, &
a_{11}^{\mathrm{SS}}&=-\frac{7}{6\pi}, &
a_{20}^{\mathrm{SS}}&=\frac{1569137}{793800\pi^2}, &
a_{21}^{\mathrm{SS}}&=-\frac{6247}{1890\pi^2}, &
a_{22}^{\mathrm{SS}}&=\frac{49}{36\pi^2}.
\end{aligned}
\end{equation}
The corrections to the propagator appear as a positive power series in $q^2$ (through $w$). This IR behavior agrees with previous results \cite{Pawlowski:2025etp,Bonanno:2021squ,Fehre:2021eob,Kher:2025rve} obtained from the graviton spectral density. In this regime, the propagator exhibits only the standard massless pole at $q^2=0$.

In the UV regime, using Eq.~(\ref{formASUV}), the propagator reads
\begin{equation}
\label{propcrcricclargep0}
G_X\left(w\rightarrow\infty\right)=\frac{1}{m_p^4}\sum_{n=1}^{\infty}\sum_{j=0}^{n-1}{\frac{b_{nj}^{\left(X\right)}}{D_X^{n+1}\left(w\right)}\ln^j{\left(\frac{23w}{12\pi}\right)}\frac{1}{w^n}},
\end{equation}
where the first coefficients $b_{nj}^{(X)}$ are
\begin{equation}
\begin{aligned}
b_{10}^{\mathrm{TT}} &= -1104\pi, \quad&
b_{20}^{\mathrm{TT}} &= -20736\pi^2, \quad&
b_{21}^{\mathrm{TT}} &= 134784\pi^2, \\
b_{10}^{\mathrm{SS}} &= -2208\pi, &
b_{20}^{\mathrm{SS}} &= 103680\pi^2, &
b_{21}^{\mathrm{SS}} &= 290304\pi^2 .
\end{aligned}
\end{equation}
and
\begin{equation}
D_{\mathrm{TT}}(w)=-206+39\ln\!\left(\frac{23w}{12\pi}\right),\qquad
D_{\mathrm{SS}}(w)=-259+186\ln\!\left(\frac{23w}{12\pi}\right).
\end{equation}
The Einstein--Hilbert contribution is exactly cancelled, so that the propagator is entirely governed by the non-local sector and exhibits an inverse-power decay in $q^2$ with logarithmic corrections. Since the functions $D_X(w)$ may vanish, the propagators develop additional poles at
\begin{equation}
w_{\mathrm{TT}}=\frac{12\pi}{23}e^{206/39},\qquad
w_{\mathrm{SS}}=\frac{12\pi}{23}e^{259/186}.
\end{equation}
corresponding to $q^2\simeq 322\,m_p^2$ and $q^2\simeq 6.57\,m_p^2$ in the TT and SS sectors, respectively. 
Similar poles in the background graviton propagator have also been reported in \cite{Knorr:2021niv}.

In the regime $D_{\mathrm{TT}}\gg1$ and $D_{\mathrm{SS}}\gg1$, i.e. far from the poles, the propagators simplify to
\begin{equation}
\label{deepUVeuc}
G_X\left(w\rightarrow\infty,D_X\to\infty\right)=\frac{1}{m_p^4}\sum_{n=1}^{+\infty}{\frac{1}{w^n}\sum_{j=1}^{+\infty}\frac{c_{nj}^{(X)}}{\ln^j{\left(\frac{23w}{12\pi}\right)}}}
\end{equation}
where the first series coefficients are 
\begin{equation}\begin{aligned}
\label{coeffdeepUVeuc}
&c_{11}^{\mathrm{TT}}=-\frac{368\pi}{13},\quad c_{12}^{TT}=-\frac{75808\pi}{507},\quad c_{21}^{TT}=\frac{1152\pi^2}{13},\quad c_{22}^{\mathrm{TT}}=\frac{155904\pi^2}{169}\\
&c_{11}^{\mathrm{SS}}=-\frac{368\pi}{31},\quad c_{12}^{\mathrm{SS}}=-\frac{47656\pi}{2883},\quad c_{21}^{\mathrm{SS}}=\frac{8064\pi^2}{961},\quad c_{22}^{\mathrm{SS}}=\frac{785472\pi^2}{29791}\ 
\end{aligned}\end{equation}
At leading order both sectors exhibit the asymptotic behavior $\sim 1/\big(q^2\ln q^2\big)$, in agreement with the Oehme--Zimmermann super-convergence relation \cite{Oehme:1990kd,Oehme:1979ai,Bonanno:2021squ}.

\begin{figure}[t]
     \centering
     \subfigure[]{
         \centering
         \includegraphics[width=0.45\textwidth]{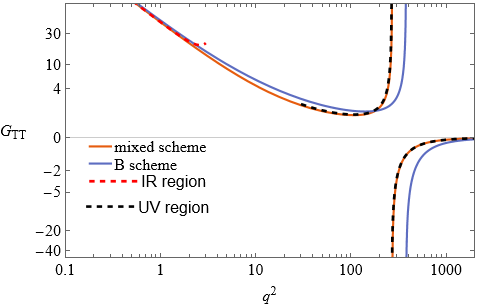}
     }
    \hspace{0.8em}
     \subfigure[]{
         \centering
         \includegraphics[width=0.45\textwidth]{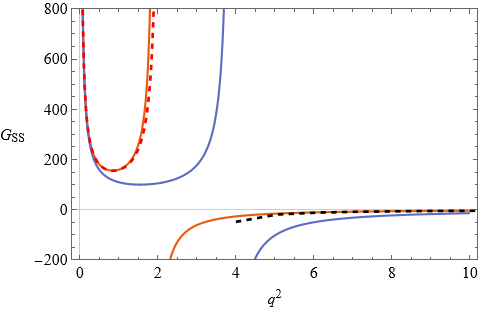}
     }
     \caption{AS Euclidean propagators in the mixed (blue) and B (beige) schemes. Dashed lines indicate the asymptotic analytical expressions in the mixed scheme. Panels (a) and (b) correspond to $G_{\mathrm{TT}}$ and $G_{\mathrm{SS}}$. The two schemes share the same functional form and differ only in the numerical values.}
     \label{plotpropASeps0}

\vspace{0.5cm}

\centering
     \subfigure[]{
         \centering
         \includegraphics[width=0.45\textwidth]{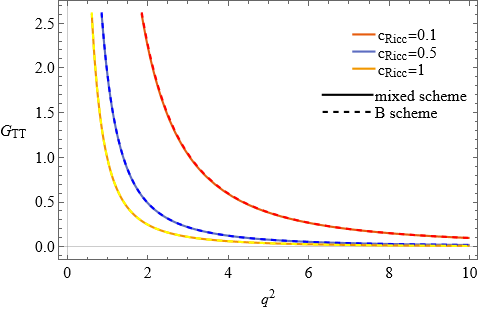}
     }
    \hspace{0.8em}
     \subfigure[]{
         \centering
         \includegraphics[width=0.45\textwidth]{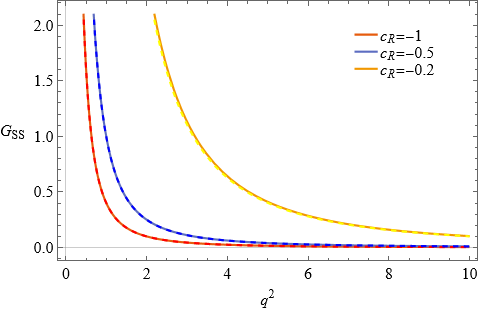}
     }
     \caption{Euclidean propagators obtained from shifted AS form factors for different values of $c_{\mathrm{R}}$ and $c_{\mathrm{Ricc}}$. Full and dashed lines correspond to the mixed and B schemes, respectively. Panels (a) and (b) show $G_{\mathrm{TT}}$ and $G_{\mathrm{SS}}$. The two schemes yield nearly indistinguishable results.}
     \label{plotpropASeps0shifted} 
     \end{figure}

Outside the asymptotic regimes, the propagators are evaluated numerically using the form factors in Fig.~\ref{plotFFASeps1}. Fig.~\ref{plotpropASeps0} shows the results, together with the corresponding asymptotic expansions. In the $G_\mathrm{TT}$ sector, the propagator exhibits a transient regime in which it decreases with increasing momentum, before turning around in the vicinity of the pole. A similar behavior occurs in the $G_{\mathrm{SS}}$ sector, although with a narrower intermediate region. For momenta larger than the pole scale, the asymptotic behavior $G_{X}\sim 1/(q^2\ln q^2)$ dominates.

A non-vanishing $c_i$, $i=R$, $Ricc$ in Eq.~(\ref{FFk=0}) modifies the UV asymptotics of the form factors, leading to $\lim_{q^2\to\infty} f(q^2)=c_i$, which in turn affects the pole structure of the propagators. The relation between $c_i$ and the pole positions $q_R^2$ and $q_{\mathrm{Ricc}}^2$ is determined by the solution of $D_X=0$:
\begin{equation}\begin{split}
&\bar c_{\mathrm{Ricc}}(q_\mathrm{Ricc}^2)=-f_\mathrm{AS}^{(\mathrm{Ricc})}(q_\mathrm{Ricc}^2)-\frac{1}{16\pi G q_\mathrm{Ricc}^2}\\
&\bar c_{\mathrm{R}}(q_\mathrm{R}^2)=-\frac{c_{\mathrm{Ricc}}}{6}-f_\mathrm{AS}^{(\mathrm{R})}(q_\mathrm{R}^2)-\frac{f^{(\mathrm{Ricc})}_\mathrm{AS}(q_\mathrm{R}^2)}{6}+\frac{1}{48\pi G q_\mathrm{R}^2}
\end{split}\end{equation}
The function $\bar c_{\mathrm{Ricc}}(q^2)$ has a maximum at $q_{\max}^2$, determined by
$f_\mathrm{AS}^{(\mathrm{Ricc})\prime}(q_{\max}^2)-\frac{1}{16\pi G\, q_{\max}^4}=0$, with $\bar c_{\mathrm{Ricc}}(q_{\max}^2)\simeq 1.4\times10^{-4}$. For $c_{\mathrm{Ricc}}>\bar c_{\mathrm{Ricc}}(q_{\max}^2)$ the pole in the $G_\mathrm{TT}$ sector disappears, leaving the propagator regular for $q^2\neq0$. In contrast, $\bar c_{\mathrm{R}}(q^2)$ is monotonic and approaches $-c_{\mathrm{Ricc}}/6$ at large $q^2$, so that the pole in the $G_{\mathrm{SS}}$ sector is removed for $c_{\mathrm{R}}<-c_{\mathrm{Ricc}}/6$.

Non-vanishing values of $c_{\mathrm{R}}$ and $c_{\mathrm{Ricc}}$ do not affect the IR scaling of the propagators, but only shift the coefficients in Eq.~(\ref{IRcoeffeucprop}). The modified coefficients read
\begin{equation}
\begin{aligned}
\label{IRcoeffshift}
\widetilde a_{10}^{\mathrm{TT}}&=a_{10}^{\mathrm{TT}}-16\pi c_{\mathrm{Ricc}}, \quad&
\widetilde a_{11}^{\mathrm{TT}}&=a_{11}^{\mathrm{TT}},\\
\widetilde a_{20}^{\mathrm{TT}}&=a_{20}^{\mathrm{TT}}-\frac{412}{45}c_{\mathrm{Ricc}}+256\pi^2c_{\mathrm{Ricc}}^2, \quad&
\widetilde a_{21}^{\mathrm{TT}}&=a_{21}^{\mathrm{TT}}-\frac{56}{3}c_{\mathrm{Ricc}}, \quad&
\widetilde a_{22}^{\mathrm{TT}}&=a_{22}^{\mathrm{TT}},\\
\widetilde a_{10}^{\mathrm{SS}}&=a_{10}^{\mathrm{SS}}+8\pi\alpha, \quad&
\widetilde a_{11}^{\mathrm{SS}}&=a_{11}^{\mathrm{SS}},\\
\widetilde a_{20}^{\mathrm{SS}}&=a_{20}^{\mathrm{SS}}+\frac{758}{45}\alpha+64\pi^2\alpha^2, \quad&
\widetilde a_{21}^{\mathrm{SS}}&=a_{21}^{\mathrm{SS}}-\frac{56}{3}\alpha, \quad&
\widetilde a_{22}^{\mathrm{SS}}&=a_{22}^{\mathrm{SS}},& 
\quad \alpha=c_{\mathrm{Ricc}}+6c_\mathrm{R}.
\end{aligned}
\end{equation}
only $a_{ii}$ is not modified. In contrast, in the UV regime non-vanishing $c_{\mathrm{R}}$ and $c_{\mathrm{Ricc}}$ modify the propagator more substantially. The Einstein--Hilbert term cancels and the leading scaling changes from $1/q^2$ to $1/q^4$, yielding
\begin{equation}
\label{propcrcricclargep}
G_X\left(q^2,w\rightarrow\infty\right)=\frac{1}{m_p^4w^2}\sum_{n=0}^{\infty}\sum_{j=0}^{n}{{\widetilde{b}}_{nj}^{\left(X\right)}\ln^j{\left(\frac{23w}{12\pi}\right)}\frac{1}{w^n}}
\end{equation}
The first coefficients of the expansion are
\begin{equation}
\begin{aligned}
\widetilde b_{00}^{\mathrm{TT}}&=\frac{1}{c_{\mathrm{Ricc}}}, \quad&
\widetilde b_{10}^{\mathrm{TT}}&=-\frac{103}{552\pi c_{\mathrm{Ricc}}^2}, \quad&
\widetilde b_{11}^{\mathrm{TT}}&=\frac{13}{368\pi c_{\mathrm{Ricc}}^2},\\
\widetilde b_{20}^{\mathrm{TT}}&=\frac{10609-5184\pi^2c_{\mathrm{Ricc}}}
{304704\pi^2c_{\mathrm{Ricc}}^3},\quad &
\widetilde b_{21}^{\mathrm{TT}}&=
\frac{13(-103+864\pi^2c_{\mathrm{Ricc}})}
{101568\pi^2c_{\mathrm{Ricc}}^3}, \quad&
\widetilde b_{22}^{\mathrm{TT}}&=
\frac{169}{135424\pi^2c_{\mathrm{Ricc}}^3},\\[3mm]
\widetilde b_{00}^{\mathrm{SS}}&=-\frac{2}{\alpha}, &
\widetilde b_{10}^{\mathrm{SS}}&=-\frac{259}{552\pi\alpha^2}, &
\widetilde b_{11}^{\mathrm{SS}}&=\frac{31}{92\pi\alpha^2},\\[2mm]
\widetilde b_{20}^{\mathrm{SS}}&=
\frac{-67081+51840\pi^2\alpha}
{609408\pi^2\alpha^3}, &
\widetilde b_{21}^{\mathrm{SS}}&=
\frac{7(1147+1728\pi^2\alpha)}
{50784\pi^2\alpha^3}, &
\widetilde b_{22}^{\mathrm{SS}}&=
-\frac{961}{16928\pi^2\alpha^3}.
\end{aligned}
\end{equation}
Representative propagators for different values of $c_{\mathrm{R}}$ and $c_{\mathrm{Ricc}}$ are shown in Fig.~\ref{plotpropASeps0shifted}. 

The propagators in the B scheme display the same qualitative structure as in the mixed scheme, with differences only at the level of numerical values (see Fig.~\ref{plotpropASeps0}). The shifted propagators are instead almost indistinguishable for all values of $c_{\mathrm{Ricc}}$ and $c_{\mathrm{R}}$ (see Fig.~\ref{plotpropASeps0shifted}).

\section{The running of the two-point function \texorpdfstring{$\Gamma_k^{(2)}[\bar g,q^2]$}{N}}\label{secEFFgamma2}
\noindent In this section we investigate how the RG flow of asymptotically safe form factors affects the scale dependence of the propagators.

\subsection{The effective dimensional Newtonian coupling}
The momentum dependence of the propagators can be parametrized in terms of effective Newton couplings:
\begin{equation}
\label{defGeff}
G_{k,\mathrm{TT}}\left(p^2\right)=\frac{16\pi G_{k,\mathrm{TT}}^{\mathrm{eff}}(q^2)}{q^2},\quad \quad G_{k,\mathrm{SS}}\left(p^2\right)=\frac{G_{k,\mathrm{SS}}^{\mathrm{eff}}(q^2)}{q^2}
\end{equation}
where
\begin{equation}
\label{Geffkgen}
G_{k,\mathrm{TT}}^{\mathrm{eff}}\left(q^2\right)=\frac{G_k}{1+16\pi G_kq^2f_k^{\left(\mathrm{Ricc}\right)}\left(q^2\right)},\quad\quad G_{k,\mathrm{SS}}^{\mathrm{eff}}\left(q^2\right)=\frac{G_k}{1+16\pi G_kq^2\left(\frac{2\left(d-1\right)}{2-d}f_k^{\left(\mathrm{R}\right)}\left(q^2\right)+\frac{f_k^{\left(\mathrm{Ricc}\right)}\left(q^2\right)}{2-d}\right)}
\end{equation}
which isolate the momentum-dependent dressing of the propagators induced by the non-local form factors. The associated anomalous dimensions are defined as
\begin{equation}\begin{split}
&\eta_k^{\mathrm{TT}}\left(q^2\right)=\frac{k\partial_kG_{k,\mathrm{TT}}^{\mathrm{eff}}\left(q^2\right)}{G_{k,\mathrm{TT}}^{\mathrm{eff}}\left(q^2\right)}=\frac{\eta_k-16\pi G_kq^2k\partial_kf_k^{\left(\mathrm{Ricc}\right)}\left(q^2\right)}{1+16\pi G_kq^2f_k^{\left(\mathrm{Ricc}\right)}\left(q^2\right)}\\
&\eta_k^{\mathrm{SS}}\left(q^2\right)=\frac{k\partial_kG_{k,\mathrm{SS}}^{\mathrm{eff}}\left(q^2\right)}{G_{k,\mathrm{SS}}^{\mathrm{eff}}\left(q^2\right)}=\frac{\eta_k-16q^2\pi G_k\left(\frac{2\left(d-1\right)}{2-d}k\partial_kf_k^{\left(\mathrm{R}\right)}\left(q^2\right)+\frac{k\partial_kf_k^{\left(\mathrm{Ricc}\right)}\left(q^2\right)}{2-d}\right)}{1+16q^2\pi G_k\left(\frac{2\left(d-1\right)}{2-d}f_k^{\left(\mathrm{R}\right)}\left(q^2\right)+\frac{f_k^{\left(\mathrm{Ricc}\right)}\left(q^2\right)}{2-d}\right)}
\end{split}\end{equation}
which characterize the RG scaling of the effective propagators. The corresponding undressed Newton coupling and anomalous dimension are given in eq.~(\ref{unimproGeta}).

The inverse of the effective couplings defines a momentum-dependent wave function renormalization:
\begin{equation}
\label{eqZ}
Z_k^{\mathrm{TT}}(q^2)=\frac{1}{16\pi G_{k,\mathrm{TT}}^{\mathrm{eff}}(q^2)},\quad\quad Z_k^{\mathrm{SS}}(q^2)=\frac{1}{16\pi G_{k,\mathrm{SS}}^{\mathrm{eff}}(q^2)}
\end{equation}
which capture the momentum-dependent corrections to the two-point function generated by the non-local form factors. 

The RG properties of $G_k^{\mathrm{eff}}(q^2)$ encode all information about the running of the propagator and the two-point function. However, the presence of additional poles in the reference propagators complicates the interpretation. To avoid this ambiguity, we restrict the analysis using the shifted form factors.

\begin{figure}[t]
    \centering
    \includegraphics[width=0.45\textwidth]{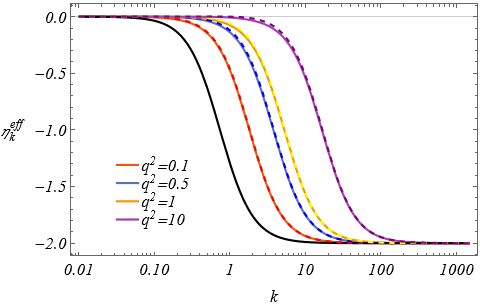}
     \caption{Effective anomalous dimensions as function of $k$ for different values of $q^2$. The full and dashed lines show $\eta_k^{\mathrm{TT}}(q^2)$ and $\eta_k^{\mathrm{SS}}(q^2)$ respectively. The black line shows the undressed case $\eta_k$ for comparison. The solutions are obtained with $c_{\mathrm{Ricc}}=1$ and $c_{\mathrm{R}}=-1/2$.}
     \label{plotetaeffk}
\end{figure}

The two anomalous dimensions share a very similar behavior. Fig.~\ref{plotetaeffk} shows $\eta_k^{\mathrm{TT}}(q^2)$ (solid lines) and $\eta_k^{\mathrm{SS}}(q^2)$ (dashed lines) for different values of $q^2$, together with the anomalous dimension $\eta_k$ (black line). Both exhibit the same crossover between the Gaussian and non-Gaussian fixed-point regimes, with only mild quantitative differences between the TT and SS sectors. This indicates an effective universality of the couplings at the graviton two-point vertex \cite{Eichhorn:2018ydy,Eichhorn:2018akn,Eichhorn:2018nda}. 

Due to the close similarity between the two sectors, in the following we focus on $G_{k,\mathrm{TT}}^{\mathrm{eff}}(q^2)$. The corresponding results for $G_{k,\mathrm{SS}}^{\mathrm{eff}}(q^2)$ are qualitatively analogous. We also restrict the analysis to the mixed scheme, since the shifted propagators in the mixed and B schemes show no significant quantitative differences within the explored range of parameters for $c_{\mathrm{R}}$ and $c_{\mathrm{Ricc}}$ (see fig. \ref{plotpropASeps0shifted}). Before starting the analysis, we stress that neither $G_{k,\mathrm{TT}}^{\mathrm{eff}}(q^2)$ nor $G_{k,\mathrm{SS}}^{\mathrm{eff}}(q^2)$ should be identified with a momentum-dependent dressing of the fundamental Newton coupling appearing in the Einstein--Hilbert action \cite{Anber:2011ut,Donoghue:2019clr,Donoghue:2024uay}. 

The asymptotics of the effective Newtonian coupling at $k=0$ can be read directly from Eqs.~(\ref{propcrcriccsmallp}) and (\ref{propcrcricclargep}). We obtain:
\begin{equation}\begin{split}
\label{IRGeffk0}
&G_{k=0}^{\mathrm{eff}}\left(w\to0\right)=\frac{1}{m_p^2}\left[1+\sum_{i=1}^{+\infty}{\sum_{j=0}^{i}{a_{ij}\left(c_{\mathrm{Ricc}}\right)\ln^j{\left(\frac{23w}{12\pi^2}\right)}}w^i}\right]\\
&G_{k=0}^{\mathrm{eff}}\left(w\rightarrow\infty\right)=\frac{1}{m_p^2}\sum_{i=1}^{+\infty}{\sum_{j=0}^{i-1}{b_{ij}\left(c_{\mathrm{Ricc}}\right)\ln^j{\left(\frac{23w}{12\pi^2}\right)}}\frac{1}{w^i}}
\end{split}\end{equation}
the IR coefficients are given in eq.(\ref{IRcoeffshift}), whereas the leading UV coefficients are 
\begin{equation}\begin{split}
&b_{10}=\frac{1}{16\pi c_{\mathrm{Ricc}}},\quad \quad b_{20}=-\frac{103}{8832\pi^2c_{\mathrm{Ricc}}^2},\quad \quad b_{21}=\frac{13}{5888\pi^2c_{\mathrm{Ricc}}^2}
\end{split}\end{equation}
In the IR, the coupling receives power-law corrections in $w$ with logarithmic contributions, while in the UV the effective Newtonian coupling is dynamically suppressed and decays as inverse powers of $w$ with logarithmic corrections.

\begin{figure}[t]
     \centering
     \subfigure[]{
         \centering
         \includegraphics[width=0.45\textwidth]{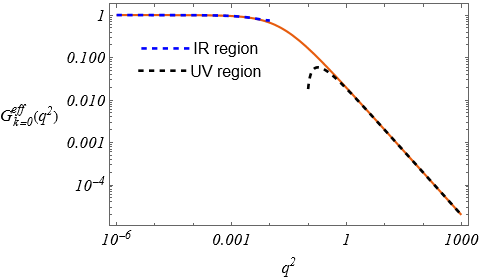}
         \label{plotGeffq}
     }
    \hspace{0.8em}
     \subfigure[]{
         \centering
         \includegraphics[width=0.45\textwidth]{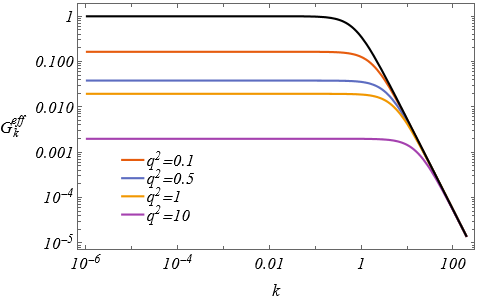}
         \label{plotGeffk}
     }
     \caption{In (a) plot of $G_{k=0}^{\mathrm{eff}}(q^2)$ and comparison with the asymptotic series (dashed lines). In (b) $G_{k=0}^{\mathrm{eff}}(q^2)$ as function of $k$ for different values of $q^2$. The black line shows $G_k$ for comparison.}
\end{figure}

Fig.~\ref{plotGeffq} shows $G_{k=0}^{\mathrm{eff}}(q^2)$ in log-log scale, together with its IR and UV asymptotic regimes. A transition region starting around $q^2 \sim m_p^2/16\pi$ separates the two asymptotic regimes, with UV scaling setting in at $q^2 \sim m_p^2$. This behavior is consistent with the results of \cite{Bonanno:2021squ}.

Fig.~\ref{plotGeffk} shows $G_k^{\mathrm{eff}}(q^2)$ as a function of $k$ for different values of $q^2$ with $m_p=1$, together with the coupling $G_k$ (black line). The running closely follows that of $G_k$, remaining approximately constant for $k \lesssim m_p$ and decreasing to zero in the UV regime. In dimensionless terms, this behavior corresponds to a smooth crossover between Gaussian and non-Gaussian fixed-point regimes, which will be analyzed in more detail in the following.

\subsection{The RG properties of the effective anomalous dimension}
Introducing the dimensionless variables 
\begin{equation}
g_k=k^{d-2}G_k,\quad\quad x=\frac{p^2}{k^2},\quad\quad F_k\left(x\right)=k^{4-d}f_k\left(p^2=xk^2\right), 
\end{equation}
one can define a dimensionless effective Newton coupling as
\begin{equation}
\label{Geffandgeff}
G_k^{\mathrm{eff}}\left(p^2=xk^2\right)\equiv k^{2-d}g_k^{\mathrm{eff}}\left(x\right),\quad\quad\ g_k^{\mathrm{eff}}\left(x\right)=\frac{g_k}{1+16\pi g_kxF_k\left(x\right)}
\end{equation}
and write the effective anomalous dimension as
\begin{equation}
\eta_k^{\mathrm{eff}}\left(x\right)\equiv\eta_k^{\mathrm{TT}}\left(p^2=xk^2\right)=\frac{\eta_k-16\pi x g_k\left[\left(d-4\right)F_k\left(x\right)-2x\partial_xF_k\left(x\right)+k\partial_kF_k\left(x\right)\right]}{1+16\pi g_kxF_k\left(x\right)}
\end{equation}
In the following we restrict to $d=4$ and make use of some results of \cite{Glaviano:2026lew}, summarized in App.~\ref{App:asymFF}. 

\begin{figure}[t]
     \centering
     \subfigure[]{
         \centering
         \includegraphics[width=0.45\textwidth]{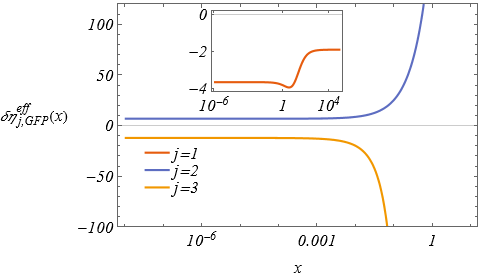}
         \label{plotetaperGFP}
     }
    \hspace{0.8em}
     \subfigure[]{
         \centering
         \includegraphics[width=0.45\textwidth]{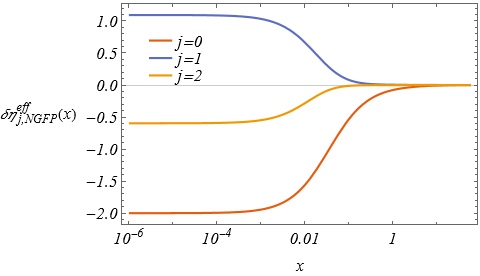}
          \label{plotetaperGNGFP}
     }
     \caption{Plots of the perturbations of $\eta_k^{\mathrm{eff}}(x)$ as function of $x$. Panel (a)  and (b) shows respectively the gaussian and non-gaussian case. In (b) $j=0$ corresponds to the dressed fixed-point $\eta_*^{\mathrm{eff}}(x)$. The perturbations are computed for $c_{\mathrm{Ricc}}=1$ and $c_{\mathrm{R}}=-1/2$.}
\end{figure}

The Gaussian regime can be analyzed by expanding $\eta_k$, $g_k$, and $F_k(x)$ in the limit $k^2 \ll m_p^2$. Using eq.~(\ref{pertGFPF}), the effective anomalous dimension takes the form
\begin{equation}\begin{split}
\label{etaeffGFP}
&\eta_{k,\mathrm{GFP}}^{\mathrm{eff}}\left(x\right)=\sum_{i=1}^{+\infty}{\delta\eta_{j,\mathrm{GFP}}^{\mathrm{eff}}\left(x\right)\left(\frac{k}{m_p}\right)^{2j}}
\end{split}\end{equation}
where the first terms are given by
\begin{equation}\begin{split}
&\delta\eta_{1,\mathrm{GFP}}^{\mathrm{eff}}\left(x\right)=-\frac{23}{2\pi}+32\pi x^2F_{\ast,\mathrm{GFP}}\left(x\right)\\
&\delta\eta_{2,\mathrm{GFP}}^{\mathrm{eff}}\left(x\right)=\frac{529}{8\pi^2}+32\pi x^2F_{1,\mathrm{GFP}}^\prime(x)-32\pi xF_{1,\mathrm{GFP}}(x)-184F_{\ast,\mathrm{GFP}}^\prime\left(x\right)-\\
&-8xF_{\ast,\mathrm{GFP}}\left(x\right)\left(64\pi^2x^2F_{\ast,\mathrm{GFP}}^\prime\left(x\right)-23\right)
\end{split}\end{equation}
the expressions for the perturbations of the form factor can be read from eq.(\ref{pertGFPF}). The asymptotic behaviors for $x\ll1$ and $x\gg1$ are given by
\begin{equation}\begin{split}
\label{etaGFxsmall}
&\delta\eta_{j,\mathrm{GFP}}^{\mathrm{eff}}\left(x\ll1\right)=\partial_k^j\eta_k|_{k=0}+a_1x+O(x^2)\\
&\delta\eta_{j,\mathrm{GFP}}^{\mathrm{eff}}\left(x\gg1\right)=b_0 x^{j-1}+O(x^{j-2})
\end{split}\end{equation}
For small $x$, the leading term reproduces the GFP regime of the undressed anomalous dimension. In contrast, for large $x$ the perturbations exhibit a power-law growth proportional to $x^{j-1}$, with the case $j=1$ being the only one where the perturbation approaches a constant. The values of the coefficients are not relevant for the present discussion. Figure~\ref{plotetaperGFP} displays the cases $j=1$ (inset, red curve), $j=2$ (blue curve) and $j=3$ (yellow curve). Deviations from the small-$x$ regime become visible around $x\sim q^2/(16\pi m_p^2)$, after which the expected $x^{j-1}$ scaling sets in.

The effective behavior around the non-trivial fixed point of $g_k$ can be determined by expanding $\eta_k$, $g_k$ and $F_k(x)$ in the limit $m_p^2\ll k^2$. Using eq.(\ref{pertNGFPF}) shows that
\begin{equation}\begin{split}
\label{etaeffNGFP}
&\eta_{k,\mathrm{NGFP}}^{\mathrm{eff}}\left(x\right)=\sum_{i=0}^{+\infty}{\delta\eta_{j,\mathrm{NGFP}}^{\mathrm{eff}}\left(x\right)\left(\frac{m_p}{k}\right)^{2j}}\ 
\end{split}\end{equation}
where the first terms of the expansion are
\begin{equation}\begin{split}
&\delta\eta_{0,\mathrm{NGFP}}^{\mathrm{eff}}\left(x\right)=\frac{2\left(-23+64\pi^2x^2F_{\ast,\mathrm{NGFP}}^\prime\left(x\right)\right)}{23+64\pi^2xF_{\ast,\mathrm{NGFP}}\left(x\right)}\\
&\delta\eta_{1,\mathrm{NGFP}}^{\mathrm{eff}}\left(x\right)=\frac{8\pi}{\left(23+64\pi^2xF_{\ast,\mathrm{NGFP}}\left(x\right)\right)^2}\Bigg(23-64\pi^2x^2F_{\ast,\mathrm{NGFP}}^\prime\left(x\right)-\\
&-32\pi x F_{1,\mathrm{NGFP}}\left(x\right)\left(-23-32\pi^2xF_{\ast,\mathrm{NGFP}}\left(x\right)+32\pi^2x^2F_{\ast,\mathrm{NGFP}}^\prime\left(x\right)\right)+\\
&+16\pi x^2\left(23+64\pi^2xF_{\ast,\mathrm{NGFP}}\left(x\right)\right)F_{1,\mathrm{NGFP}}^\prime\left(x\right)\Bigg)
\end{split}\end{equation}
here $\delta\eta_{0,\mathrm{NGFP}}^{\mathrm{eff}}\left(x\right)$ defines the effective fixed-point solution and the higher-order terms describe the corresponding perturbations. The asymptotic behavior of the coefficients is
\begin{equation}\begin{split}
\label{etaNGFPx}
&\delta\eta_{j,\mathrm{NGFP}}^{\mathrm{eff}}\left(x\ll1\right)=\partial_{m_p^2}^j\eta_{k}|_{m_p^2=0}+a_1x+O(x^2)\\
&\delta\eta_{j,\mathrm{NGFP}}^{\mathrm{eff}}\left(x\gg1\right)=b_0 x^{-1}+O(x^{-2})
\end{split}\end{equation}
Thus, the NGFP regime of the undressed anomalous dimension is recovered at small $x$, while at large $x$ the effective anomalous dimension decays as inverse powers of $x$. Similar asymptotic behavior has been observed also for the fluctuation anomalous dimension in \cite{Christiansen:2014raa,Christiansen:2015rva,Knorr:2021niv, Denz:2016qks}.

Fig. \ref{plotetaperGNGFP} shows the $x$ dependence of effective non-trivial fixed-point behavior (red line) and its perturbations with $j=1$ (blue line) and $j=2$ (yellow line). As for the gaussian case, the deviation from the small $x$ regime starts around $x\sim q^2/16\pi m_p^2$. 

Outside the fixed-point regime for $F_k$, the previous expansions are no longer valid. In particular, the asymptotic behaviors for small and large $x$ are no longer reliable. In this case, the full expansion for small and large $x$ is required. From eq.(\ref{aeFASsmallk}) for $x\ll1$ the solution reads
\begin{equation}\begin{split}
\label{IRefateff}
\eta_k^{\mathrm{eff}}\left(x\ll1\right)=\eta_k+\sum_{n=1}^{+\infty}{\sum_{m=0}^{n}{a_{nm}\left(k\right)\ln^m{\left(\frac{x}{3}\right)}}x^n}
\end{split}\end{equation}
this is valid for all $k$. The first coefficients are given by
\begin{equation}\begin{split}
&a_{10}=\frac{7\eta_k}{115}+\left(-\frac{1009}{10350}+\frac{32\pi^2}{23}c_{\mathrm{Ricc}}+\frac{7}{138}\ln{\left(-\frac{2}{\eta_k}\right)}\right)\eta_k^2,\quad\quad a_{11}=-\frac{28}{345}\eta_k^2\\
&a_{20}=-\frac{17\eta_k}{966}+\eta_k^2\left(\frac{224\pi^2}{2645}c_{\mathrm{Ricc}}+\frac{49\ln{\left(-\frac{2}{\eta_k}\right)}}{7935\left(\eta_k+2\right)}+\frac{81292}{4165875}\right)\\
&+\eta_k^3\left(\frac{122089}{26780625}-\frac{23888\pi^2}{119025}c_{\mathrm{Ricc}}\frac{1024\pi^4c_{\mathrm{Ricc}}^2}{529}+\frac{\left(2822400\pi^2c_{\mathrm{Ricc}}-251029\right)\ln{\left(-\frac{2}{\eta_k}\right)}}{9998100\left(\eta_k+2\right)}+\frac{49\ln^2{\left(-\frac{2}{\eta_k}\right)}}{19044}\right)+\\
&+\eta_k^4\left(\frac{224\pi^2}{1587\left(\eta_k+2\right)}c_{\mathrm{Ricc}}-\frac{10451}{1428300\left(\eta_k+2\right)}\right)\ln{\left(-\frac{2}{\eta_k}\right)}\\
&a_{21}=-\frac{196\eta_k^2}{39675}+\eta_k^3\left(\frac{20902}{1785375}-\frac{1792\pi^2}{7935}c_{\mathrm{Ricc}}-\frac{196\ln{\left(-\frac{2}{\eta_k}\right)}}{23805}\right),\quad \quad a_{22}=\frac{784\eta_k^3}{119025}
\end{split}\end{equation}
At $x=0$ the undressed anomalous dimension is recovered for all $k$, while the corrections are given by smooth power series modulated by logarithmic terms.

In the large $x$ regime, using eq.(\ref{aeFASlargek}) we find
\begin{equation}
\label{UVefateff}
\eta_k^{\mathrm{eff}}\left(x\gg1\right)=\sum_{n=1}^{+\infty}{\sum_{m=0}^{n-1}{b_{nm}\left(k\right)\ln^m{\left(\frac{x}{3}\right)}}\frac{1}{x^n}}
\end{equation}
where the first coefficient is given by
\begin{equation}\begin{split}
&b_{10}=-\frac{3}{8c_{\mathrm{Ricc}}\pi^2}\left(1+\frac{13\eta_k}{24}\right)\\
&b_{20}=-\frac{\frac{103}{128\pi^4}+\frac{39\ln{\left(-\frac{2}{\eta_k}\right)}}{256\pi^4}}{c_{\mathrm{Ricc}}^2\eta_k}-\frac{99}{16\pi^2c_{\mathrm{Ricc}}}-\frac{\frac{325\ln{\left(-\frac{2}{\eta_k}\right)}}{2048\pi^4}+\frac{2593}{3072\pi^4}}{c_{\mathrm{Ricc}}^2}+\frac{\eta_k}{c_{\mathrm{Ricc}}}\left(\frac{117}{32\pi^2}-\frac{\frac{2717}{12288\pi^4}+\frac{169\ln{\left(-\frac{2}{\eta_k}\right)}}{4096\pi^4}}{c_{\mathrm{Ricc}}}\right)\\
&b_{21}=\frac{1}{c_{\mathrm{Ricc}}^2}\left(\frac{39}{256\pi^4\eta_k}+\frac{325}{2048\pi^4}+\frac{169\eta_k}{4096\pi^4}\right)
\end{split}\end{equation}
Eq.(\ref{UVefateff}) is valid away from the Gaussian regime and shows that $\eta_k^{\mathrm{eff}}$ decays as inverse powers of $x$ so that the scaling $x^{j-1}$ does not extend beyond the Gaussian regime. The decay at large $x$ implies momentum locality and absence of quadratic divergences in the flow of $\Gamma_k^{(2)}(x)$  \cite{Christiansen:2015rva,Christiansen:2014raa}. 

\subsection{The RG properties of the effective Newtonian coupling}
The effective dimensionless Newtonian coupling satisfies a flow equation given by
\begin{equation}
\label{efffloweq}
k\partial_kg_k^{\mathrm{eff}}\left(x\right)=\left(d-2+\eta_k^{\mathrm{eff}}\left(x\right)\right)g_k^{\mathrm{eff}}\left(x\right)+2x\partial_xg_k^{\mathrm{eff}}\left(x\right)
\end{equation}
which is a partial differential equation. Its solution must be consistent with Eq.~(\ref{Geffkgen}) and reduce to $g_k$ at $x=0$. This yields
\begin{equation}
\label{solgeff}
g_k^{\mathrm{eff}}(x)
= g_\ast\,\exp\!\left[\int_{-\infty}^{k}\frac{d-2+\eta_v^{\mathrm{eff}}\!\left(\frac{k^2 x}{v^2}\right)}{v}\,dv\right].
\end{equation}
where $g_*$ is the undressed NGFP value. This expression makes explicit how the effective anomalous dimension controls the scale and momentum dependence of $g_k^{\mathrm{eff}}(x)$ in Eq.~(\ref{Geffandgeff}).

The effective flow equation in Eq.~(\ref{efffloweq}) allows one to interpret the running of $G_k^{\mathrm{eff}}(q^2)$ as governed by a genuine RG scale-dependent coupling. In this framework, the asymptotic regimes in $k/m_p$ of the effective anomalous dimension directly correspond to the fixed-point structure of $g_k^{\mathrm{eff}}(x)$. This provides a clear explanation of what Fig. \ref{plotGeffk} shows.

\begin{figure}[t]
     \centering
     \subfigure[]{
         \centering
         \includegraphics[width=0.45\textwidth]{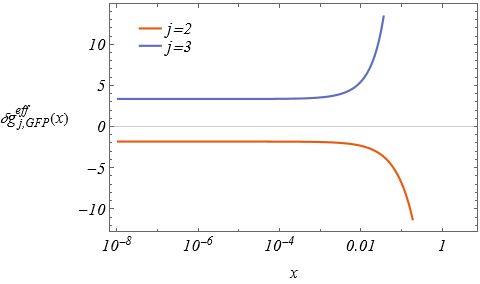}
         \label{plotgeffGFP}
     }
    \hspace{0.8em}
     \subfigure[]{
         \centering
         \includegraphics[width=0.45\textwidth]{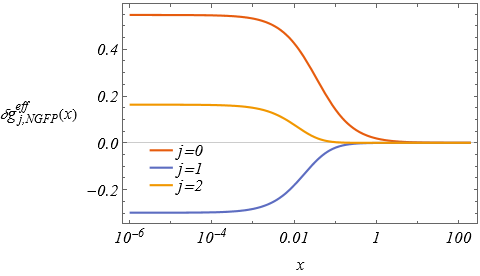}
     }
     \caption{Plots of the perturbations of $g_k^{\mathrm{eff}}(x)$ as function of $x$. Panel (a)  and (b) shows respectively the gaussian and non-gaussian case. In (b) $j=0$ corresponds to the non-trivial dressed fixed-point $g_*$. The couplings are computed for $c_{\mathrm{Ricc}}=1$ and $c_{\mathrm{R}}=-1/2$.}
     \label{plotgeffNGFP}

\vspace{0.5cm}

    \centering
    \includegraphics[width=0.45\textwidth]{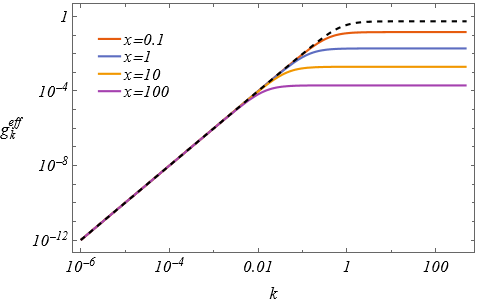}
     \caption{Plot of $g_k^{\mathrm{eff}}(x)$ as a function of $k$ for different values of $x$ and comparison with the undressed case $g_k$ (black line). The behavior is the same as $g_k$. The couplings are computed for $c_{\mathrm{Ricc}}=1$ and $c_{\mathrm{R}}=-1/2$.}
     \label{plotgeffk}
\end{figure}

In the dressed Gaussian regime, the coupling admits the expansion
\begin{equation}
\label{geffGFP}
g_k^{\mathrm{eff}}\left(x\right)=\frac{k^2}{m_p^2}+\sum_{j=2}^{+\infty}{\delta g_{j,\mathrm{GFP}}^{\mathrm{eff}}\left(x\right)\left(\frac{k}{m_p}\right)^{2j}}
\end{equation}
showing that the Gaussian fixed point remains IR attractive and UV repulsive. The asymptotics in $x$ follows the same structure as Eq.~(\ref{etaGFxsmall}):
\begin{equation}\begin{split}
\label{geffGFPxsmall}
&\delta g_{j,\mathrm{GFP}}^{\mathrm{eff}}\left(x\ll1\right)=\partial_k^jg_k|_{k=0}+a_1x+O(x^2)\\
&\delta g_{j,\mathrm{GFP}}^{\mathrm{eff}}\left(x\gg1\right)=b_0 x^{j-1}+O(x^{j-2})
\end{split}\end{equation}
with small-$x$ regularity and large-$x$ growth. Fig. \ref{plotgeffGFP} shows the perturbations for $j=2$ (beige line) and $j=3$ (blue line). The onset of deviations from the undressed coupling occurs around $x\sim q^2/16\pi m_p^2$, beyond which the coupling exhibits the expected large $x$ power-law scaling.

In the dressed non-gaussian fixed point regime the solution takes the form
\begin{equation}
\label{geffNGFP}
g_k^{\mathrm{eff}}\left(x\right)=\sum_{j=0}^{+\infty}{\delta g_{j,\mathrm{NGFP}}^{\mathrm{eff}}\left(x\right)\left(\frac{m_p}{k}\right)^{2j}}
\end{equation}
where the leading term corresponds to the dressed NGFP solution, which is UV attractive and IR repulsive. The asympotics in $x$ of this expansion follow the same structure of eq.(\ref{etaNGFPx}):
\begin{equation}\begin{split}
\label{gNGFPx}
&\delta g_{j,\mathrm{NGFP}}^{\mathrm{eff}}\left(x\ll1\right)=\partial_{m_p^2}^jg_{k}|_{m_p^2=0}+a_1x+O(x^2)\\
&\delta g_{j,\mathrm{NGFP}}^{\mathrm{eff}}\left(x\gg1\right)=b_0 x^{-(j+1)}+O(x^{-3}),\quad\quad j\ge1
\end{split}\end{equation}
at small $x$ one recovers the undressed NGFP regime with regular power-law corrections, while the large-$x$ behavior is suppressed by inverse powers of $x$. This is consistent with the results of \cite{Bonanno:2021squ}. Fig. \ref{plotgeffNGFP} shows the fixed-point (beige line), and the perturbations for $j=1$ (blue) and $j=2$ (yellow). The deviation from the undressed case occurs again above $x\sim q^2/16\pi m_p^2$.

Outside the fixed-point regimes, the full IR expansion follows the same structure of eq.(\ref{IRefateff}):
\begin{equation}\begin{split}
\label{serieIRgeff}
g_k^{\mathrm{eff}}\left(x\right)=g_k+\sum_{n=1}^{+\infty}{\sum_{m=0}^{n}{a_{nm}\left(k\right)\ln^m{\left(\frac{x}{3}\right)}}x^n}
\end{split}\end{equation}
The first terms are
\begin{equation}\begin{split}
&a_{10}=g_k^2\left(\frac{121}{225\pi}-16\pi c_{\mathrm{Ricc}}-\frac{7\ln{\left(\frac{4\pi}{23g_k}\right)}}{12\pi}\right),\quad\quad a_{11}=\frac{14g_k^2}{15\pi}\\
&a_{20}=-\frac{17g_k^2}{168\pi}+g_k^3\left(\frac{\left(121-3600\pi^2c_{\mathrm{Ricc}}\right)^2}{50625\pi^2}+\frac{\left(2822400\pi^2c_{\mathrm{Ricc}}-230449\right)\ln{\left(\frac{4\pi}{23g_k}\right)}}{37800\pi\left(4\pi-23g_k\right)}+\frac{49\ln{\left(\frac{4\pi}{23g_k}\right)}}{144\pi^2}\right)-\\
&-\frac{161\left(3600\pi^2c_{\mathrm{Ricc}}-121\right)g_k^4\ln{\left(\frac{4\pi}{23g_k}\right)}}{1350\pi^2\left(4\pi-23g_k\right)}\\
&a_{21}=\left(\frac{3388}{3375\pi^2}-\frac{448c_{\mathrm{Ricc}}}{15}-\frac{49\ln{\left(\frac{4\pi}{23g_k}\right)}}{45\pi^2}\right)g_k^3,\quad \quad a_{22}=\frac{196g_k^3}{225\pi^2}
\end{split}\end{equation}
at $x=0$ the undressed solution is always recovered and the corrections at $g_k$ are encoded in an expansion in powers of $x$ modulated by logarithmic terms. 

At large $x$ the solution follows the structure of eq.(\ref{UVefateff}) and is given by
\begin{equation}
\label{serieUVgeff}
g_k^{\mathrm{eff}}\left(x\gg1\right)=\sum_{n=1}^{+\infty}\sum_{m=0}^{n-1}{b_{nm}\ln^m{\left(\frac{x}{3}\right)}\frac{1}{x^n}}
\end{equation}
The first terms are
\begin{equation}\begin{split}
&b_{10}=\frac{1}{16\pi x c_{\mathrm{Ricc}}},\quad \quad a_{11}=\frac{14g_k^2}{15\pi}\\
&b_{20}=-\frac{103}{8832\pi^2c_{\mathrm{Ricc}}^2g_k}+\frac{209}{3072\pi^3c_{\mathrm{Ricc}}^2},\quad \quad b_{21}=\frac{13}{1024\pi^3c_{\mathrm{Ricc}}^2}-\frac{13}{5888\pi^2c_{\mathrm{Ricc}}^2g_k}\\
\end{split}\end{equation}
in this regime, the effective coupling scales as inverse powers of $x$. As for the effective anomalous dimension, the gaussian scaling does not extend outside the Gaussian regime.

Fig. \ref{plotgeffk} shows the flow of $g_k^{\mathrm{eff}}(x)$ as function of $k$ for different values of $x$ and the comparison with the undressed $g_k$ (black dashed line). The dressed runnings follow the same features of $g_k$, interpolating smoothly between a gaussian and a non-gaussian fixed point regime.

\section{Analytical continuations to the Minkowski spacetime}\label{secAnalcont}
\noindent In this section, we discuss the analytic continuation of the form factors to Minkowski spacetime.

The continuation of the form factors is performed via a Wick rotation, mapping the Euclidean momentum to the Minkowskian one, $q_E^2 \rightarrow -q_M^2$. In terms of $w$ in eq. (\ref{varadim}), this implies
\begin{equation}
w \rightarrow -w_M, 
\end{equation}
Unlike the Euclidean case, the Minkowskian variable 
$w_M$ is not restricted only to positive values.

Analytically continuing the IR and UV expansions in Eqs.~(\ref{formASIR}) and (\ref{formASUV}) shows that the logarithmic terms generate
\begin{equation}
\ln\!\left(\frac{23w}{12\pi}\right)=\ln\!\left(-\frac{23w_M}{12\pi}\right)=\ln\!\left(\frac{529w_M^2}{144\pi^2}\right)+i\pi\,\theta(w_M)+2i\pi n,\quad \quad n\in \mathbb{Z}
\end{equation}
implying that for $w_M>0$ the form factors develop an imaginary part. The asymptotic structure of the expansions is preserved, while the coefficients in Eqs.~(\ref{formASIR}) and (\ref{formASUV}) are modified as
\begin{equation}
\tilde a_j^{(i)} = (-1)^j[a_j^{(i)} + i\,b_j^{(i)}\,\left(\theta(w_M)+2n\right)],
\qquad
\tilde b_j^{(i)} = \frac{(-1)^j}{2}\,b_j^{(i)}.
\end{equation}
This reflects a branch cut structure in Minkowski space induced by the logarithmic terms.

Outside the asymptotic regimes, the analytic continuation has to be constructed numerically. We employ interpolating functions that reproduce the Euclidean form factors in in Fig. \ref{plotFFASeps1} over the full momentum range, match the analytic asymptotic expansions, and implement the continuation $q_E^2 \to -q_M^2$ in a smooth way. In practice, the construction is non-trivial, as the interpolation must avoid introducing spurious zeros or poles upon continuation. Polynomial or rational ansätze do not in general satisfy all these requirements simultaneously, so we restrict the analysis to $q_M^2>0$, where such artifacts are minimized. For simplicity we set also $n=0$.

\begin{figure}[t]
     \centering
     \subfigure[]{
         \centering
         \includegraphics[width=0.45\textwidth]{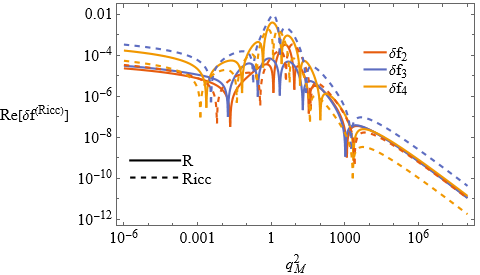}
     }
    \hspace{0.8em}
     \subfigure[]{
         \centering
         \includegraphics[width=0.45\textwidth]{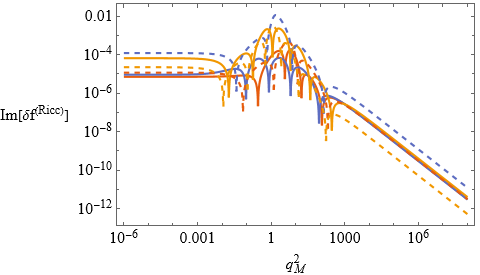}
     }
     \caption{Plot of $\delta f=|f_\mathrm{AC}^{(1)}(w_M)-f_\mathrm{AC}^{(i)}(w_M)|$ for $i=2,3,4$. Panels (a) and (b) show the real and imaginary parts, respectively. Solid and dashed lines correspond to the Ricci scalar and Ricci tensor sectors, respectively. The analytic continuation is stable under the choice of interpolating function.}
     \label{plotstab}
\end{figure}

To assess the robustness of the continuation, we consider different admissible analytic fits:
\begin{equation}
\label{fitASAC}
\begin{aligned}
f_\mathrm{AC}^{(1)}(w_M)
&= \frac{\sum_{j=0}^{1}(a_j+c_j\ln w_M)w_M^j}{1+d_1 w_M+d_2 w_M^2},
&
f_\mathrm{AC}^{(2)}(w_M)
&= \frac{\sum_{j=0}^{3}(a_j+c_j \ln w_M)w_M^j}{1+d_4w_M^4}
\\
f_\mathrm{AC}^{(3)}(w_M)
&= \frac{\sum_{j=0}^{3}(a_j+c_j\ln w_M)w_M^j}{(1+d_2w_M^2)(1+d_1w_M+d_4w_M^2)},
&
\quad f_\mathrm{AC}^{(4)}(w_M)
&= \frac{\sum_{j=0}^{5}(a_j+c_j \ln w_M)w_M^j}{(1+d_2w_M^2)(1+d_4w_M^4)}
\end{aligned}
\end{equation}
which reproduce the asymptotics without introducing spurious poles or zeros. The coefficients are obtained by fitting the data in Fig.~\ref{plotFFASeps1}. Taking $f_\mathrm{AC}^{(1)}$ as reference, we plot $\delta f_i=|f_\mathrm{AC}^{(1)}-f_\mathrm{AC}^{(i)}|$ for $i=2,3,4$. As shown in Fig.~\ref{plotstab}, the differences satisfy $\delta f_i<10^{-3}$ over the full momentum range, indicating that the analytic continuation is stable under the choice of admissible interpolating functions and that the intermediate-momentum region is under numerical control.

\begin{figure}[p]
     \centering
     \subfigure[]{
         \centering
         \includegraphics[width=0.45\textwidth]{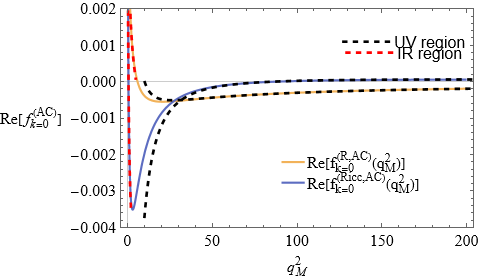}
     }
    \hspace{0.8em}
     \subfigure[]{
         \centering
         \includegraphics[width=0.45\textwidth]{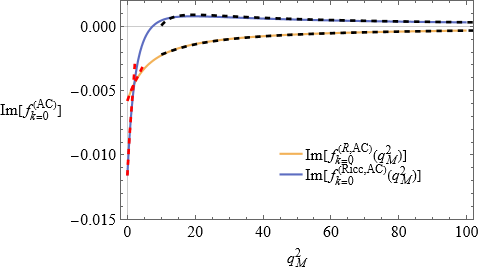}
     }
     \caption{Reference analytically continued numerical form factors at $k=0$ in the mixed scheme and comparison with the analytical asymptotic expressions. The plot shows the real and imaginary part respectively in (a) and (b).}
     \label{plotFFASACeps0}

\vspace{0.5cm}

     \centering
     \subfigure[]{
         \centering
         \includegraphics[width=0.45\textwidth]{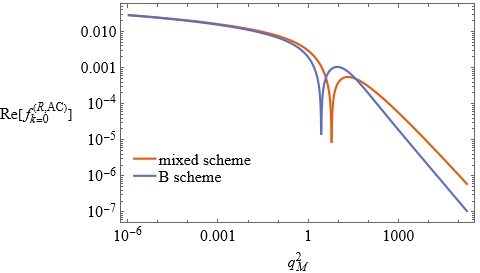}
     }
    \hspace{0.8em}
     \subfigure[]{
         \centering
         \includegraphics[width=0.45\textwidth]{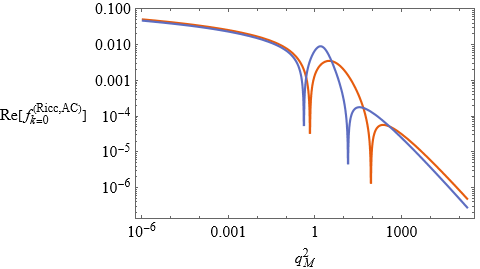}
     }
     \caption{Real part of the reference asymptotically safe form factors in the B scheme, compared with the mixed scheme results of Fig.~\ref{plotFFASACeps0}. Panels (a) and (b) show $\mathrm{Re}[f_{k=0,\mathrm{AC}}^{(\mathrm{R})}(q^2)]$ and $\mathrm{Re}[f_{k=0,\mathrm{AC}}^{(\mathrm{Ricc})}(q^2)]$, respectively. As in the Euclidean case, the functional structure is scheme independent.}
    \label{plotFFReeps1}

\vspace{0.5cm}

     \centering
     \subfigure[]{
         \centering
         \includegraphics[width=0.45\textwidth]{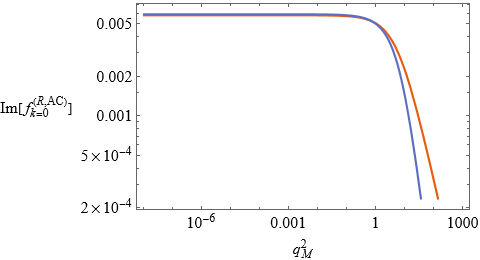}
     }
    \hspace{0.8em}
     \subfigure[]{
         \centering
         \includegraphics[width=0.45\textwidth]{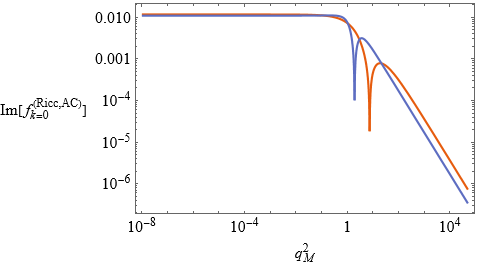}
     }
     \caption{Imaginary part of the reference asymptotically safe form factors in the B scheme, compared with the mixed scheme results of Fig.~\ref{plotFFASACeps0}. Panels (a) and (b) show $\mathrm{Im}[f_{k=0,\mathrm{AC}}^{(\mathrm{R})}(q^2)]$ and $\mathrm{Im}[f_{k=0,\mathrm{AC}}^{(\mathrm{Ricc})}(q^2)]$, respectively. The functional structure is scheme independent.}
     \label{plotFFImeps1}
     \end{figure}

Fig.~\ref{plotFFASACeps0} shows the resulting analytically continued form factors together with their asymptotic expansions. The real parts exhibit a transition region between IR and UV regimes, where both form factors develop a negative minimum. In the UV, $\mathrm{Re}[f_\mathrm{AC}^{(\mathrm{R})}]$ approaches zero from below, while $\mathrm{Re}[f_\mathrm{AC}^{(\mathrm{Ricc})}]$ changes sign and approaches zero from above. The imaginary parts remain negative for $f_\mathrm{AC}^{(\mathrm{R})}$ and develop a positive intermediate maximum for $f_\mathrm{AC}^{(\mathrm{Ricc})}$.

The B-scheme results can also be continued using Eq.~(\ref{fitASAC}). Figures~\ref{plotFFReeps1} and \ref{plotFFImeps1} compare the two schemes, showing the same qualitative behavior and differing only in numerical values.

\section{The Minkowskian asymptotically safe propagator}\label{secminkprop}
\noindent In this section we discuss the graviton propagator in Minkowski spacetime using the analytically continued form factors. 

It is convenient to rewrite the propagator as
\begin{equation}\begin{split}
\label{splitprop}
&\frac{1}{\frac{1}{16\pi G}q^2\left(1+16\pi G q^2\Sigma(q^2)\right)}
\equiv G_1(q^2)+G_2(q^2)\,,\\
&G_1(q^2)=\frac{1}{\frac{1}{16\pi G}q^2}, \qquad
G_2(q^2)=-\frac{256\pi^2 G^2\,\Sigma(q^2)}{1+16\pi G q^2\Sigma(q^2)},
\end{split}\end{equation}
where $\Sigma(p^2)$ collects the form-factor contributions appearing in Eq.~(\ref{propTTSS}). This decomposition isolates in a transparent way the corrections induced by the non-local terms.

The Euclidean and Minkowskian propagators share the same formal structure, although their analytic properties differ. To treat pole and branch-cut contributions, we adopt the Feynman prescription $q_M^2\to q_M^2+i\epsilon$ and use the decomposition in Eq.~(\ref{splitprop}). Separating real and imaginary parts, the Einstein--Hilbert contribution reads
\begin{equation}
\label{propG1}
G_1(q_M^2)=\frac{1}{\frac{1}{16\pi G}\left(q_M^2+i\epsilon\right)}
= P.V.\!\left[\frac{1}{\frac{q_M^2}{16\pi G}}\right]
- i16G\pi^2\,\delta(q_M^2),
\end{equation}
where $P.V.$ denotes the Cauchy principal value. For the analytically continued form-factor contribution $G_2(q_M^2)$, writing $\Sigma(\bar q_M^2)=A(q_M^2,\epsilon)+iB(q_M^2,\epsilon)$, $\bar q_M^2=q_M^2+i\epsilon$, one obtains
\begin{equation}\begin{split}
\label{minkpropformgen}
&G_2(\bar q_M^2)=-\frac{256G^2\pi^2\left[A+16G\pi q_M^2\left(A^2+B^2\right)\right]}{\left[1+16G\pi\left(Aq_M^2-B\epsilon\right)\right]^2+\left[16G\pi\left(Bq_M^2+A\epsilon\right)\right]^2}+\\
&+i\frac{256G^2\pi^2\left(-B+16G\pi\epsilon\left(A^2+B^2\right)\right)}{\left[1+16G\pi\left(Aq_M^2-B\epsilon\right)\right]^2+\left[16G\pi\left(Bq_M^2+A\epsilon\right)\right]^2}.
\end{split}\end{equation}
In the limit $\epsilon\to0^+$, if $B\neq 0$, the denominators do not develop real poles within the class of asymptotically safe solutions considered here, and the $i\epsilon$ prescription is needed only to fix the branch of the logarithmic functions. The form-factor contribution is therefore fully regular on the real axis, with a non-vanishing imaginary part arising from branch-cut structures.

\subsection{The properties of the propagators}
In the following analysis of the propagator, the $i\epsilon$ prescription is left implicit and the limit $\epsilon\to0^+$ is understood, and omitted for notational simplicity.

In the IR regime, from eq.(\ref{propcrcriccsmallp}), the analytically continued propagators are given by 
\begin{equation}
\label{ACIRserieprop}
G_X^{\mathrm{AC}}\left(q_M^2,w_M\rightarrow0\right)=\frac{16\pi}{m_p^2q_M^2}\left[1+\sum_{n=1}^{+\infty}\sum_{j=0}^{n}{a_{nj}^{\left(X\right)}\ln^j{\left(\frac{23w_M}{12\pi}\right)}w_M^n}\right]
\end{equation}
where $X=\mathrm{TT}$, $\mathrm{SS}$. The first terms of these series are given by
\begin{equation}
\begin{alignedat}{5}
a_{10}^{\mathrm{TT}}&=\frac{103}{360\pi}+\frac{7i}{12}, 
&\qquad a_{11}^{\mathrm{TT}}&=\frac{7}{12\pi},
&\qquad a_{20}^{\mathrm{TT}}&=-\frac{49}{144}+\frac{988441}{1587600\pi^2}
+\frac{37211i}{30240\pi},
\\[1mm]
a_{21}^{\mathrm{TT}}&=-\frac{17023}{30240\pi^2}+\frac{49i}{72\pi},
&\qquad a_{22}^{\mathrm{TT}}&=\frac{49}{9\pi},
\\[2mm]
a_{10}^{SS}&=\frac{379}{360\pi}-\frac{7i}{6},
&\qquad a_{11}^{\mathrm{SS}}&=-\frac{7}{6\pi},
&\qquad a_{20}^{\mathrm{SS}}&=-\frac{49}{36}+\frac{761861}{3175200\pi^2}
-\frac{6247i}{1890\pi},
\\[1mm]
a_{21}^{\mathrm{SS}}&=-\frac{6077}{3780\pi^2}+\frac{49i}{18\pi},
&\qquad a_{22}^{\mathrm{SS}}&=\frac{49}{36\pi^2}.
\end{alignedat}
\end{equation}
The corrections to the propagators scale as the Euclidean result but now also an imaginary part is present. In this regime only the classical pole at $q_M^2=0$ is present.

In the UV, from eq.(\ref{propcrcricclargep0}) we find
\begin{equation}
G_X\left(w_M\rightarrow\infty\right)=\frac{1}{m_p^4}\left[\frac{a_{10}^{\left(X\right)}+a_{11}^{\left(X\right)}\ln{\left(\frac{23w_M}{12}\right)}}{D_{X}\left(w_M\right)w_M}+\sum_{n=2}^{+\infty}\sum_{j=0}^{n+1}\frac{a_{nj}^{\left(X\right)}\ln^j{\left(\frac{23w_M}{12}\right)}}{D_{X}^n\left(w_M\right)w_M^n}\right]
\end{equation}
where
\begin{equation}\begin{split}
&D_{\mathrm{TT}}\left(w_M\right)=4624+1521\pi^2-5304\ln{\left(\frac{23w_M}{12\pi}\right)}+1521\ln^2{\left(\frac{23w_M}{12\pi}\right)},\\
&D_{\mathrm{SS}}\left(w_M\right)=289+34596\pi^2+6324\ln{\left(\frac{23w_M}{12\pi}\right)}+34596\ln^2{\left(\frac{23w_M}{12\pi}\right)}.
\end{split}\end{equation}
with the first coefficients given by
\begin{equation}\begin{split}
&a_{10}^{\mathrm{TT}}=-75072\pi-43056i\pi^2,\quad a_{11}^{\mathrm{TT}}=43056\pi,\\
&a_{20}^{\mathrm{TT}}=-20736\pi^2\left(4624+32955\pi^2\right)-134784i\pi^3\left(1521\pi^2-3808\right)\\
&a_{21}^{\mathrm{TT}}=134784\pi^2\left(5440+1521\pi^2\right)+63078912i\pi^3,\\
&a_{22}^{\mathrm{TT}}=-746433792\pi^2-205006464i\pi^3,\quad a_{23}^{\mathrm{TT}}=205006464\pi^2
\end{split}\end{equation}
and
\begin{equation}\begin{split}
&a_{10}^{\mathrm{SS}}=37536\pi-410688i\pi^2,\quad a_{11}^{\mathrm{SS}}=410688\pi,\\
&a_{20}^{\mathrm{SS}}=-20736\pi^2\left(84444\pi^2-1445\right)-41472i\pi^3\left(13787+242172\pi^2\right)\\
&a_{21}^{\mathrm{SS}}=41472\pi^2\left(17833+242172\pi^2\right)-7173826560i\pi^3,\\
&a_{22}^{\mathrm{SS}}=5422795776\pi^2-10043357184i\pi^3,\quad a_{23}^{\mathrm{SS}}=10043357184\pi^2
\end{split}\end{equation}
These series exhibit the same qualitative behavior as in the Euclidean case. The denominators have a zero at
\begin{equation}
w_{\mathrm{TT}}=\frac{12\pi}{23}\exp\!\left(\frac{5304+3042i\pi}{3042}\right),\quad \quad 
w_{\mathrm{SS}}=\frac{12\pi}{23}\exp\!\left(\frac{-6324+69192i\pi}{69192}\right).
\end{equation}
Unlike the Euclidean case, these zeros lie in the complex plane, suggesting resonance-like structures. However, within the class of background propagators considered here, and given the freedom in the boundary conditions of the fixed-point form factors, we do not attribute them a direct physical interpretation.

In the regime $D_{\mathrm{TT}}\gg1$ and $D_{\mathrm{SS}}\gg1$, the propagators exhibit the same asymptotic structure as Eq.~(\ref{deepUVeuc}):
\begin{equation}
G_X\left(w_M\rightarrow\infty,D_X\to\infty\right)=\frac{1}{m_p^4}\sum_{n=1}^{+\infty}{\frac{1}{w_M^n}\sum_{j=1}^{+\infty}\frac{b_{nj}}{\ln^j{\left(\frac{23w_M}{12\pi}\right)}}}
\end{equation}
with the first coefficients given by
\begin{equation}\begin{aligned}
&b_{11}^{\mathrm{TT}}=\frac{368\pi}{13},\quad\quad b_{12}^{\mathrm{TT}}=\frac{25024\pi}{507}-\frac{368i\pi^2}{13},\quad\quad b_{21}^{\mathrm{TT}}=\frac{1152\pi^2}{13},\quad \quad b_{22}^{\mathrm{TT}}=\frac{3840\pi^2}{13}-\frac{1152i\pi^3}{13}\\
&b_{11}^{\mathrm{SS}}=\frac{368\pi}{31},\quad\quad b_{12}^{\mathrm{SS}}=-\frac{3128\pi}{2883}-\frac{368i\pi^2}{31},\quad\quad  b_{21}^{\mathrm{SS}}=\frac{8064\pi^2}{961},\quad\quad b_{22}^{\mathrm{SS}}=\frac{43584\pi^2}{29791}-\frac{8064i\pi^3}{961}\ 
\end{aligned}\end{equation}
The leading coefficients coincide with those of Eq.~(\ref{coeffdeepUVeuc}). The real part scales as $1/(q_M^2\ln q_M^2)$, while the imaginary part is suppressed by one additional logarithm, scaling as $1/(q_M^2\ln^2 q_M^2)$. This is consistent with the analysis of the background propagators in \cite{Bonanno:2021squ}.

\begin{figure}[t]
\vspace{0.5cm}

     \centering
     \subfigure[]{
         \centering
         \includegraphics[width=0.45\textwidth]{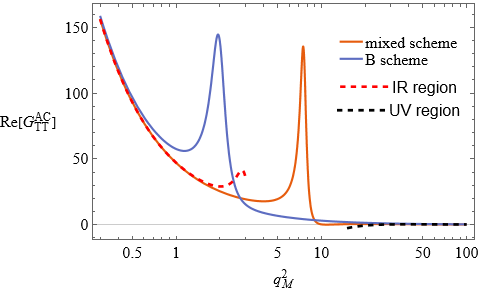}
     }
    \hspace{0.8em}
     \subfigure[]{
         \centering
         \includegraphics[width=0.45\textwidth]{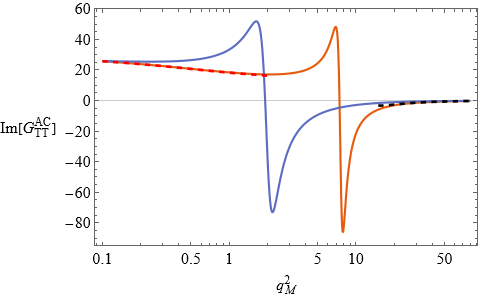}
     }
     \caption{Analytically continued $G_{\mathrm{TT}}$. Panels (a) and (b) show real and imaginary parts. The propagator has the same functional form in both schemes, differing only in numerical values. The imaginary part is negative in the UV. The spectral density is $\rho_{\mathrm{TT}}^{\mathrm{cut}}=\mathrm{Im}[G_{\mathrm{TT}}]/\pi$.}
    \label{plotpropTTACeps0}

\vspace{0.5cm}

     \centering
     \subfigure[]{
         \centering
         \includegraphics[width=0.45\textwidth]{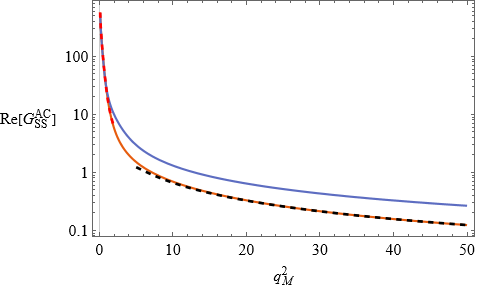}
     }
    \hspace{0.8em}
     \subfigure[]{
         \centering
         \includegraphics[width=0.45\textwidth]{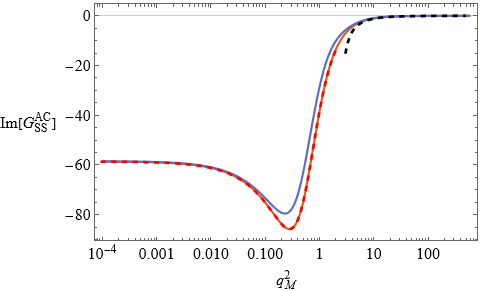}
     }
     \caption{Analytically continued $G_{\mathrm{SS}}$. Panels (a) and (b) show the real and imaginary parts, respectively. The propagator exhibits the same functional form in both schemes, differing only in numerical values. The real part is positive, while the imaginary part is negative. The spectral density is $\rho_{\mathrm{SS}}^{\mathrm{cut}}=\mathrm{Im}[G_{\mathrm{SS}}]/\pi$.}
     \label{plotpropSSACeps0}
     \end{figure}

Outside the asymptotic regimes, the analytic continuation is evaluated numerically using the interpolating function of Eq.~(\ref{fitASAC}) into the propagators. The resulting $G_{\mathrm{TT}}^{\mathrm{AC}}(q_M^2)$ and $G_{\mathrm{SS}}^{\mathrm{AC}}(q_M^2)$, shown in Figs.~\ref{plotpropTTACeps0} and \ref{plotpropSSACeps0}, interpolate between the IR and UV asymptotic behaviors, with the transition occurring around the Planck scale. In both sectors, the transition region is characterized by a local maximum before the UV regime is reached. The real parts remain positive for all momenta. In contrast, $\mathrm{Im}[G_{\mathrm{TT}}^{\mathrm{AC}}]$ changes sign between the IR and UV regimes, whereas $\mathrm{Im}[G_{\mathrm{SS}}^{\mathrm{AC}}]$ remains negative throughout.

The corresponding results in the B scheme are shown in Figs.~\ref{plotpropTTACeps0} and \ref{plotpropSSACeps0} as blue curves. Both real and imaginary parts display the same qualitative behavior as in the mixed scheme, with differences arising only in numerical values. For $\mathrm{Im}[G_{\mathrm{TT}}^{\mathrm{AC}}]$, the sign change occurs in the deep UV at $q_M^2 \sim 3 m_p^2$. 

Shifts of the form factors by the constants $c_{\mathrm{R}}$ and $c_{\mathrm{Ricc}}$, as in the Euclidean case, do not affect the IR behavior but modify the coefficients in Eq.~(\ref{ACIRserieprop}):
\begin{equation}
\begin{aligned}
\widetilde{a}_{10}^{\mathrm{TT}} &= a_{10}^{\mathrm{TT}}-16\pi c_{\mathrm{Ricc}},
&\qquad
\widetilde{a}_{11}^{\mathrm{TT}} &= a_{11}^{\mathrm{TT}},
&\qquad
\widetilde{a}_{20}^{\mathrm{TT}} &= a_{20}^{\mathrm{TT}}
+\left(-\frac{412}{45}-\frac{56 i\pi}{3}\right)c_{\mathrm{Ricc}}
+256\pi^2 c_{\mathrm{Ricc}}^2,
\\[1ex]
\widetilde{a}_{21}^{\mathrm{TT}} &= a_{21}^{\mathrm{TT}}-\frac{56}{3}c_{\mathrm{Ricc}},
&\qquad
\widetilde{a}_{22}^{\mathrm{TT}} &= a_{22}^{\mathrm{TT}},
\\[1.5ex]
\widetilde{a}_{10}^{\mathrm{SS}} &= a_{10}^{\mathrm{SS}}+8\pi\alpha,
&\qquad
\widetilde{a}_{11}^{\mathrm{SS}} &= a_{11}^{\mathrm{SS}},
&\qquad
\widetilde{a}_{20}^{\mathrm{SS}} &= a_{20}^{\mathrm{SS}}
+\left(\frac{758}{45}-\frac{56 i\pi}{3}\right)\alpha
+64\pi^2\alpha^2,
\\[1ex]
\widetilde{a}_{21}^{\mathrm{SS}} &= a_{21}^{\mathrm{SS}}-\frac{56}{3}\alpha,
&\qquad
\widetilde{a}_{22}^{\mathrm{SS}} &= a_{22}^{\mathrm{SS}},
&\qquad
\alpha &= c_{\mathrm{Ricc}}+6c_{\mathrm{R}}.
\end{aligned}
\end{equation}
As in the Euclidean case, the UV asymptotics take the form of Eq.~(\ref{propcrcricclargepAC}):
\begin{equation}
\label{propcrcricclargepAC}
G_X\left(q^2,w_M\rightarrow\infty\right)=\frac{1}{m_p^4w_M^2}\sum_{n=0}^{\infty}\sum_{j=0}^{n}{{\widetilde{b}}_{nj}^{\left(X\right)}\ln^j{\left(\frac{23w_M}{12\pi}\right)}\frac{1}{w_M^n}}
\end{equation}
where 
\begin{equation}
\begin{aligned}
\widetilde{b}_{00}^{\mathrm{TT}} &= \frac{1}{c_{\mathrm{Ricc}}},
&\quad
\widetilde{b}_{10}^{\mathrm{TT}} &= \frac{-39i+\frac{68}{\pi}}{1104c_{\mathrm{Ricc}}^2},
&\quad
\widetilde{b}_{11}^{\mathrm{TT}} &= -\frac{13}{368\pi c_{\mathrm{Ricc}}^2},
\\[1ex]
\widetilde{b}_{20}^{\mathrm{TT}}
&=
\frac{9i\left(2i+13\pi\right)}{1058c_{\mathrm{Ricc}}^2}-\frac{\left(68i+39\pi\right)^2}{1218816c_{\mathrm{Ricc}}^3\pi^2},
&\quad
\widetilde{b}_{21}^{\mathrm{TT}}
&=
\frac{117}{1058c_{\mathrm{Ricc}}^2}
+\frac{13i\left(68i+39\pi\right)}
{203136\pi^2c_{\mathrm{Ricc}}^3},
&\quad
\widetilde{b}_{22}^{\mathrm{TT}}
&=
\frac{169}{135424\pi^2c_{\mathrm{Ricc}}^3},
\\[1ex]
\widetilde{b}_{00}^{\mathrm{SS}} &= \frac{2}{\alpha},
&\quad
\widetilde{b}_{10}^{\mathrm{SS}} &= -\frac{17+186i\pi}{552\alpha^2\pi},
&\quad
\widetilde{b}_{11}^{\mathrm{SS}} &= -\frac{31}{92\alpha^2\pi},
\\[1ex]
\widetilde{b}_{20}^{\mathrm{SS}}
&=
\frac{\left(17i-186\pi\right)^2}
{609408\pi^2\alpha^3}
+\frac{9i\left(14\pi-5i\right)}
{529\alpha^2},
&\quad
\widetilde{b}_{21}^{\mathrm{SS}}
&=
\frac{126}{529\alpha^2}
-\frac{527+5766i\pi}
{50784\pi^2\alpha^3},
&\quad
\widetilde{b}_{22}^{\mathrm{SS}}
&=
-\frac{961}{16928\pi^2\alpha^3}.
\end{aligned}
\end{equation}
The inclusion of $c_{\mathrm{R}}$ and $c_{\mathrm{Ricc}}$ removes the complex poles discussed above and does not generate additional real poles.

\begin{figure}[t]
     \centering
     \subfigure[]{
         \centering
         \includegraphics[width=0.45\textwidth]{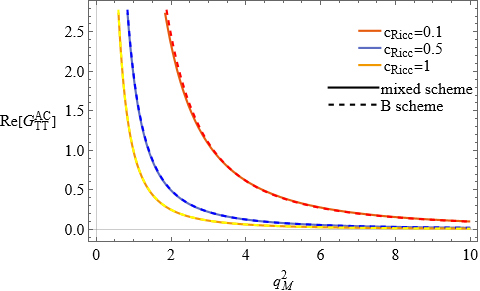}
     }
    \hspace{0.8em}
     \subfigure[]{
         \centering
         \includegraphics[width=0.45\textwidth]{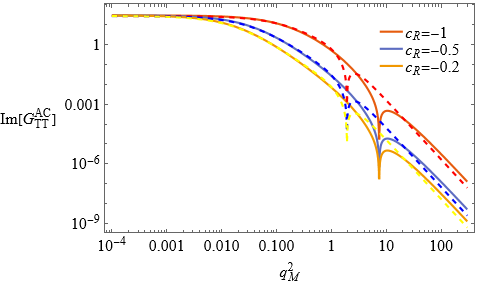}
     }
     \caption{Analytically continued shifted $G_{\mathrm{TT}}$. Panels (a) and (b) show the real and imaginary parts, respectively. The two schemes yield nearly indistinguishable results.}
     \label{plottotTTACS}

\vspace{0.5cm}

     \subfigure[]{
         \centering
         \includegraphics[width=0.45\textwidth]{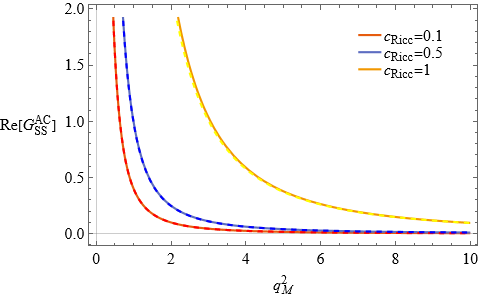}
     }
    \hspace{0.8em}
     \subfigure[]{
         \centering
         \includegraphics[width=0.45\textwidth]{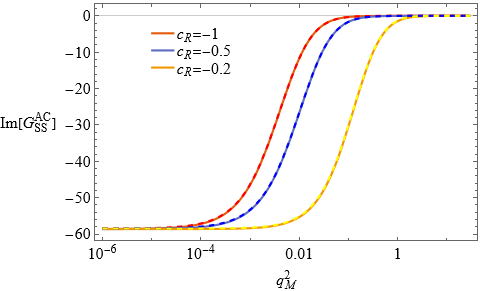}
     }
     \caption{Analytically continued shifted $G_{\mathrm{SS}}$. Panels (a) and (b) show the real and imaginary parts, respectively. The two schemes yield nearly indistinguishable results.}
     \label{plottotSSACS}
\end{figure}

Figure~\ref{plottotTTACS} and \ref{plottotSSACS} show the shifted propagators $G_{\mathrm{TT}}^{\mathrm{AC}}(q_M^2)$ and $G_{\mathrm{SS}}^{\mathrm{AC}}(q_M^2)$ for different values of $c_{\mathrm{R}}$ and $c_{\mathrm{Ricc}}$. Full and dashed lines correspond to the mixed and B schemes, respectively. The real parts closely follow their Euclidean counterparts. As in the reference case, $\mathrm{Im}[G_{\mathrm{TT}}^{\mathrm{AC}}(q_M^2)]$ is positive in the IR and negative in the UV, whereas $\mathrm{Im}[G_{\mathrm{SS}}^{\mathrm{AC}}(q_M^2)]$ remains negative for all momenta. The two schemes yield nearly indistinguishable results.

\subsection{The Källén–Lehmann representation and its interpretation}
Due to the imaginary part generated by the logarithmic terms of the form factors after analytic continuation, the propagator does not develop additional real poles beyond the massless one at $q_M^2=0$. This has important consequences for the interpretation of the Källén–Lehmann representation \cite{KL1,Kallen:1952zz,Bonanno:2021squ,Fehre:2021eob}. This is given by  
\begin{equation}
G_{\mu\nu\rho\sigma}(q_M^2)=\int_{0}^{+\infty}\frac{\rho_{\mu\nu\rho\sigma}(\mu^2)}{q_M^2-\mu^2+i\epsilon}\,d\mu^2 \, ,
\end{equation}
where the spectral density can be decomposed as
\begin{equation}
\rho_{\mu\nu\rho\sigma}(\mu^2)=\rho_{\mathrm{TT}}(\mu^2)P_{2,\mu\nu\rho\sigma}+\rho_{\mathrm{SS}}(\mu^2)P_{S,\mu\nu\rho\sigma}\,,
\end{equation}
Formally, $\rho_{\mu\nu\rho\sigma}(\mu^2)$ is related to the discontinuity of the propagator across the branch cut,
\begin{equation}
\rho_{\mu\nu\rho\sigma}(\mu^2)=-\frac{1}{\pi}\lim_{\epsilon\to0^+}\mathrm{Im}[G_{\mu\nu\rho\sigma}(\mu^2+i\epsilon)]\,.
\end{equation}
Using the decomposition eq.~(\ref{splitprop}), together with the expressions in eqs.~(\ref{propG1}) and (\ref{minkpropformgen}), the spectral density becomes
\begin{equation}\begin{split}
&\rho_{\mu\nu\rho\sigma}(\mu^2)=\rho^{\mathrm{EH}}_{\mu\nu\rho\sigma}(\mu^2)+\rho^{\mathrm{cut}}_{\mu\nu\rho\sigma}(\mu^2)\\
&\rho^{\mathrm{EH}}_{\mu\nu\rho\sigma}(\mu^2)=16\pi G\delta(\mu^2)(P_{2,\mu\nu\rho\sigma}+P_{S,\mu\nu\rho\sigma}),\quad\quad \rho^{\mathrm{cut}}_{\mu\nu\rho\sigma}(\mu^2)=\frac{1}{\pi}\mathrm{Im}[G_{2,\mu\nu\rho\sigma}(\mu^2)]\,.
\end{split}\end{equation}
Up to the factor $\pi^{-1}$, $\rho^{\mathrm{cut}}_{\mu\nu\rho\sigma}$ coincides with the imaginary parts shown in Figs.~\ref{plotpropTTACeps0} and \ref{plotpropSSACeps0}, which is not positive definite, in agreement with previous analyses of background propagators \cite{Bonanno:2021squ}.

The non-positivity of the spectral function does not imply a violation of unitarity, but rather reflects the fact that, due to the non-local momentum dependence and the absence of additional real poles, the propagator does not admit a standard Stieltjes representation with positive measure \cite{SchillingSongVondracek+2012,Briscese:2024tvc}. Consequently, the associated spectral density is generally not unique and cannot be interpreted as a probability distribution. This feature is intrinsic to the non-local structure of $G_2(q_M^2)$ in Eq.~(\ref{splitprop}) and, in the absence of additional real poles, is independent of the distinction between background and fluctuation propagators. In the presence of non-local form factors, the relevant indicator of unitarity is the pole structure rather than the requirement of a positive definite spectral density. Since $G_2(q_M^2)$ does not develop additional poles with wrong-sign residues, no ghost-like degrees of freedom arise within this approximation and the resulting infinite-derivative theory remains ghost-free.

\subsection{The dressed Minkowskian Newtonian coupling}
The analytical continuation can also be performed on the dressed Newtonian coupling in eq.~(\ref{Geffkgen}) at $k=0$, describing in this way the dressed properties of the two-point function in the Minkowski spacetime. 

Using the IR expansion in eq.~(\ref{IRGeffk0}), we obtain
\begin{equation}
\label{IRserieGeffAC}
G_{k=0}^{\left(eff,AC\right)}\left(w_M\to0\right)=\frac{1}{m_p^2}\left[1+\sum_{n=1}^{+\infty}{\sum_{m=0}^{n}{a_{nm}\left(c_{\mathrm{Ricc}}\right)\ln^m{\left(\frac{23w_M}{12\pi}\right)}}w_M^n}\right]
\end{equation}
The first series coefficients are
\begin{equation}\begin{split}
&a_{10}=\frac{103}{360\pi}-16\pi c_{\mathrm{Ricc}}+\frac{7i}{12},\quad\quad a_{11}=\frac{7}{12\pi}\\
&a_{20}=-\frac{49}{144}+\frac{988441}{1587600\pi^2}-\frac{412c_{\mathrm{Ricc}}}{45}+256\pi^2c_{\mathrm{Ricc}}^2+i\left(-\frac{17023}{30240\pi}-\frac{56\pi c_{\mathrm{Ricc}}}{3}\right)\\
&a_{21}=-\frac{17023}{30240\pi^2}-\frac{56c_{\mathrm{Ricc}}}{3}+\frac{49i}{72\pi},\quad\quad a_{22}=\frac{49}{144\pi^2}
\end{split}\end{equation}
This reproduces the structure of Eq.~(\ref{IRGeffk0}), with IR corrections to the Plank mass organized as a double expansion in $w_M$ and logarithms.

In the UV regime, using eq.~(\ref{IRGeffk0}), the solution reads
\begin{equation}
\label{UVserieGeffAC}
G_{k=0}^{\left(eff,AC\right)}\left(w_M\to\infty\right)=\frac{1}{m_p^2}\left[\sum_{n=1}^{+\infty}{\sum_{m=0}^{n-1}{a_{nm}\left(c_{\mathrm{Ricc}}\right)\ln^m{\left(\frac{23w_M}{12\pi}\right)}}\frac{1}{w_M^n}}\right]
\end{equation}
The first series coefficients are 
\begin{equation}\begin{split}
&a_{10}=\frac{1}{16\pi c_{\mathrm{Ricc}}},\quad\quad a_{11}=\frac{7}{12\pi},\quad\quad a_{20}=\frac{17}{4416\pi^2c_{\mathrm{Ricc}}^2}-\frac{13i}{5888\pi c_{\mathrm{Ricc}}^2},\quad \quad a_{21}=-\frac{13}{5888\pi^2c_{\mathrm{Ricc}}^2}
\end{split}\end{equation}
This matches the structure of eq.~(\ref{IRGeffk0}) and shows a decay in inverse powers of $w_M$.

\begin{figure}[t]
     \centering
     \subfigure[]{
         \centering
         \includegraphics[width=0.45\textwidth]{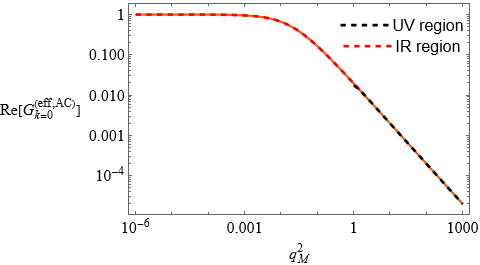}
         \label{plotGeffACTTRe}
     }
    \hspace{0.8em}
     \subfigure[]{
         \centering
         \includegraphics[width=0.45\textwidth]{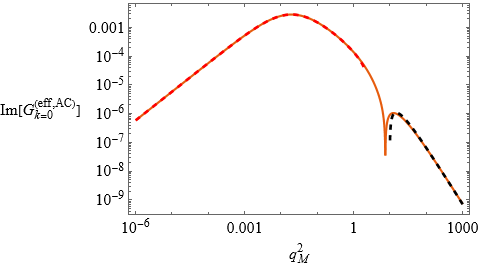}
         \label{plotGeffACTTIm}
     }
     \caption{Analytically continued $G_{k=0}^{(\mathrm{eff,AC})}(q_M^2)$. Panels (a) and (b) show $\mathrm{Re}\,G_{k=0}^{(\mathrm{eff,AC})}(q_M^2)$ and $\mathrm{Im}\,G_{k=0}^{(\mathrm{eff,AC})}(q_M^2)$, respectively. The red and black dashed lines correspond to eqs.~(\ref{IRserieGeffAC}) and (\ref{UVserieGeffAC}), respectively.}
\end{figure}

Fig.~\ref{plotGeffACTTRe} and Fig.~\ref{plotGeffACTTIm} show the real and imaginary parts of the analytically continued effective coupling across the full momentum range, together with the IR and UV asymptotic expansions. The real part follows the same qualitative behavior as in the Euclidean case, while the imaginary part develops a maximum at intermediate scales, changes sign, and approaches the UV regime from below.

\section{The corrections to the Newtonian potential}\label{secNewtpot}
\noindent In this section, we describe how the presence of the form factors in the propagator affects the Newtonian potential.

To derive the corrections to the Newtonian potential, we consider static perturbations $h_{\mu\nu}(\mathbf{x})$
and introduce a localized matter source at $r=0$, with energy-momentum tensor
\begin{equation}
T_{\mu\nu}=\rho(\mathbf{x})\,\delta_\mu^0\delta_\nu^0\,,
\end{equation}
For simplicity, we consider a point-particle density $\rho(\mathbf{x})=m$. The Newtonian potential is then obtained from the Fourier transform of the non-relativistic scattering amplitude,
\begin{equation}
\label{fourtrans}
V(r)=-\int\frac{d^3q}{(2\pi)^3}e^{-iq\cdot r}M(p^2)
=-\frac{m}{4}\int\frac{d^3p}{(2\pi)^3}e^{-iq\cdot r}G_{0000}(q^2)\,,
\end{equation}
where
\begin{equation}
\label{G0000}
G_{0000}(q^2)=\frac{2}{3}G_{\mathrm{TT}}(q^2)+\frac{1}{3}G_{\mathrm{SS}}(q^2)\,.
\end{equation}
The integral is well defined since the integrand is regular at $q\to0$ and sufficiently suppressed for $q\to\infty$, ensuring convergence.

The Euclidean propagators associated with the reference asymptotically safe form factors exhibit additional real poles beyond $q^2=0$, leading to an oscillatory contribution in the corresponding Newtonian potential that is difficult to interpret physically. We therefore restrict the analysis to the shifted form factors, for which no additional Euclidean poles arise provided that $c_{\mathrm{Ricc}}>1.4\times10^{-5}$ and $c_{\mathrm{R}}< -c_{\mathrm{Ricc}}/6$.

The Newtonian potential cannot be computed in closed form, but its asymptotic behaviour in $r$ can be derived analytically without evaluating the full momentum integral; the derivation is reported in Appendix~\ref{appNewpot}.

In the large-$r$ regime, the Newtonian potential takes the form
\begin{equation}\begin{split}
\label{newpotIRexp}
V_{IR}(r,c_{\mathrm{R}},c_{\mathrm{Ricc}})=-\frac{mG}{r}\left(1+\sum_{n=2}^{+\infty}\sum_{m=0}^{n-1}
a_{nm}(c_{\mathrm{R}},c_{\mathrm{Ricc}})\frac{\ln^m\!\left(\frac{12}{23}\pi r^2m_p^2\right)}{(m_pr)^{2n}}\right),
\end{split}\end{equation}
with coefficients determined by Eq.~(\ref{newpotIR}). The first ones are
\begin{equation}\begin{split}
&a_{20}=-448c_{\mathrm{R}}-224c_{\mathrm{Ricc}}-\frac{98\gamma}{3\pi^2}+\frac{4747}{84\pi^2},\quad a_{21}=-\frac{49}{3\pi^2}\\
&\resizebox{1.05\hsize}{!}{$a_{30}=-\frac{103693}{60\pi^3}+\frac{1715}{18\pi}+\gamma\left(\frac{62720c_{\mathrm{R}}}{\pi}-\frac{7066}{3\pi^3}\right)+\frac{3430\gamma^2}{3\pi^3}-\frac{758752c_{\mathrm{R}}}{7\pi}+\frac{90704c_{\mathrm{Ricc}}}{7\pi}+53760\pi\left(6c_{\mathrm{R}}-c_{\mathrm{Ricc}}\right)\left(2c_{\mathrm{R}}+c_{\mathrm{Ricc}}\right)$},\\
&a_{31}=\frac{31360c_{\mathrm{R}}}{\pi}+\frac{3430\gamma}{3\pi^3}-\frac{3533}{3\pi^3},\quad a_{32}=\frac{1715}{6\pi^3}.
\end{split}\end{equation}
where $\gamma$ is the Euler constant. The IR corrections generate higher powers of $1/r$ together with logarithmic corrections. The would-be leading $1/r^3$ term vanishes due to cancellations among the logarithmic contributions in the IR expansion of $G_{0000}(q^2)$.

\begin{figure}[t]
     \centering
     \subfigure[]{
         \centering
         \includegraphics[width=0.45\textwidth]{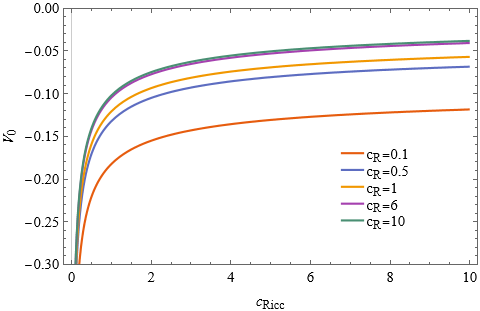}
         \label{plotnewtpotV0}
     }
    \hspace{0.8em}
     \subfigure[]{
         \centering
         \includegraphics[width=0.45\textwidth]{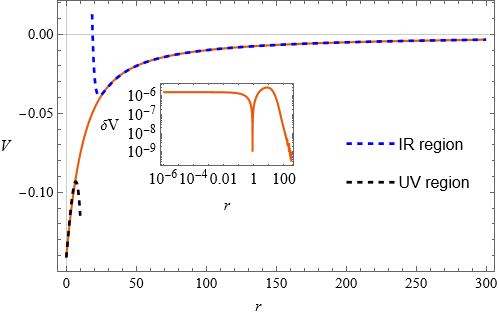}
         \label{plotnewtpot}
     }
     \caption{In (a) values of $V(r=0)$ as function of $c_{\mathrm{Ricc}}$ for different values of $c_{\mathrm{R}}$. In (b) the Newtonian potential obtained with $(c_{\mathrm{Ricc}},c_{\mathrm{R}})=(1,-0.5)$ and comparison with the corresponding analytical asymptotic expressions (dashed lines). In the inset of (b) the absolute difference $\delta V=|V_{\mathrm{B}}(r)-V_{\mathrm{MS}}(r)|$ is shown. The potential is almost scheme independent.}
\end{figure}

In the small-$r$ regime, the potential reads
\begin{equation}\begin{split}
\label{newpotUVexp}
V_{UV}(r)=\frac{m}{m_p}\left(V_0(c_{\mathrm{R}},c_{\mathrm{Ricc}})+\sum_{n=0}^{+\infty}\sum_{m=0}^{n}
b_{nm}(c_{\mathrm{R}},c_{\mathrm{Ricc}})\ln^m\!\left(\frac{12}{23}\pi r^2m_p^2\right)(m_pr)^{2n+1}\right),
\end{split}\end{equation}
with coefficients given in Eq.~(\ref{newpotUV}). The first ones are
\begin{equation}\begin{split}
&b_{00}=\frac{c_{\mathrm{R}}}{8\pi c_{\mathrm{Ricc}}(6c_{\mathrm{R}}+c_{\mathrm{Ricc}})},\\
&b_{10}=\frac{116c_{\mathrm{R}}c_{\mathrm{Ricc}}+348c_{\mathrm{R}}^2-105c_{\mathrm{Ricc}}^2}{141312\pi^2c_{\mathrm{Ricc}}^2(6c_{\mathrm{R}}+c_{\mathrm{Ricc}})^2}
+\frac{\gamma(52c_{\mathrm{R}}c_{\mathrm{Ricc}}+156c_{\mathrm{R}}^2+25c_{\mathrm{Ricc}}^2)}{35328\pi^2c_{\mathrm{Ricc}}^2(6c_{\mathrm{R}}+c_{\mathrm{Ricc}})^2},\\
&b_{11}=\frac{52c_{\mathrm{R}}c_{\mathrm{Ricc}}+156c_{\mathrm{R}}^2+25c_{\mathrm{Ricc}}^2}{70656\pi^2c_{\mathrm{Ricc}}^2(6c_{\mathrm{R}}+c_{\mathrm{Ricc}})^2}.
\end{split}\end{equation}
The UV contributions remove the $1/r$ divergence, yielding a regular expansion of the potential near the origin. The constant term $V_0$, obtained from Eq.~(\ref{c0newtpot}) and computed numerically, is shown in Fig.~\ref{plotnewtpotV0} as a function of $c_{\mathrm{Ricc}}$ for different values of $c_{\mathrm{R}}$ (with $m_p=1$). In all cases $V_0<0$, it approaches zero for $c_{\mathrm{Ricc}}\to\infty$, while increasing $c_{\mathrm{R}}$ shifts it to smaller values. 

The regularity of the potential at $r=0$ is consistent with expectations from gravitational theories with non-local form factors \cite{Biswas:2011ar,Biswas:2005qr,Modesto:2011kw,Frolov:2015usa,Giacchini:2016xns}. In the asymptotically safe scenario, from the results for the effective Newtonian coupling in sec. \ref{secEFFgamma2}, this behavior is associated with gravitational antiscreening induced by the UV-attractive fixed point, which suppresses the effective Newton coupling at high energies  \cite{Platania:2023srt,Falls:2010he,Platania:2025imw,Bonanno:2017pkg}. 

Outside the asymptotic regimes, the potential is computed numerically. Figure~\ref{plotnewtpot} shows a representative example with $c_{\mathrm{R}}=-0.5$, $c_{\mathrm{Ricc}}=1$, together with the asymptotic expansion. The potential is negative for all $r$ and grows in magnitude as $r\to 0$, reaching its minimum $V(0)=V_0\simeq-0.14$. 

The Newtonian potential obtained from the asymptotically safe form factors in the B scheme does not differ significantly from the corresponding mixed scheme case. The absolute difference $\delta V=|V_{\mathrm{B}}(r)-V_{\mathrm{MS}}(r)|$ is shown in the inset of Fig.~\ref{plotnewtpot} and is always smaller than $10^{-6}$. This indicates that the potential is essentially insensitive to the RG scheme choice.

\section{Conclusions}\label{secconc}
\noindent In this work we used the flow of the background asymptotically safe form factors derived in \cite{Glaviano:2026lew} to analyze their implications for the graviton propagator. Deriving propagators in the presence of non-local form factors is technically challenging, as it requires the inversion of the Hessian discussed in App.~\ref{appderbeta}, which can presently be carried out only around flat spacetime. The resulting Euclidean propagators at $k=0$ exhibit a non-trivial momentum dependence: in the IR, the classical $1/q^2$ behavior receives corrections organized as positive powers of $q^2$ accompanied by logarithmic terms, eq.~(\ref{propcrcriccsmallp}), while in the UV the Einstein--Hilbert contribution cancels and the reference propagator scales as $1/[q^2\ln(q^2)]$, eq.~(\ref{propcrcricclargep0}). Consequently, the asymptotic scaling is fully governed by the non-local sector. Away from the asymptotic regimes, additional real poles can appear, but they can be removed by exploiting the freedom in the boundary conditions used to solve the fixed-point equation for the form factors. This leaves the IR behavior unchanged, while the UV scaling is effectively governed by a $1/q^4$ behavior.

The requirement of avoiding additional poles in the background Euclidean propagators selects the shifted form factors as the preferred non-local structures. These depend on two a priori unconstrained boundary parameters, $c_{\mathrm{R}}$ and $c_{\mathrm{Ricc}}$. Within the present approximation, the absence of extra poles restricts them to the region $c_{\mathrm{Ricc}}\geq 1.4\times10^{-4}$ and $c_{\mathrm{R}}<-c_{\mathrm{Ricc}}/6$. Including fluctuation form factors may further constrain this parameter space through additional consistency conditions. In this setup, the deep-UV effective action is characterized by $\mathcal{L}(q^2\gg m_p^2)\sim R+c_{\mathrm{R}}R^2+c_{\mathrm{Ricc}}R_{\mu\nu}R^{\mu\nu}+O(m_p^2/q^2)$. Whether this structure persists in more general settings remains an open question for future investigations.

The flow of the AS form factors characterizes the running of the graviton two-point function. The TT and SS sectors can be parametrized in terms of effective Newton couplings $G_{\mathrm{TT}}^{\mathrm{eff}}(q^2)$ and $G_{\mathrm{SS}}^{\mathrm{eff}}(q^2)$. At fixed dimensionless momentum $x=q^2/k^2$, the corresponding dimensionless couplings $g_k^{\mathrm{eff}}(x)$ interpolate between the  Gaussian and non-Gaussian fixed-point regimes as the undressed Newton coupling $g_k$. As functions of $x$, they smoothly connect the undressed value at $x=0$ to an inverse-power decay at large $x$, except in the Gaussian regime where a power-law growth is observed. The effective anomalous dimensions display the same pattern. Altogether, at the background level, these results are consistent with previous studies \cite{article,Christiansen:2012rx,Christiansen:2014raa,Christiansen:2015rva,Denz:2016qks,Bonanno:2021squ} and indicate that the asymptotically safe fixed-point structure remains stable in the presence of non-local form factors. At $k=0$, the momentum dependence of $G_{k=0}^{\mathrm{eff}}(q^2)$ reproduces the qualitative running of $G_k$ under the identification $k^2\leftrightarrow q^2$, in agreement with \cite{Bonanno:2021squ}.

The form factors can be analytically continued to Minkowski spacetime via a Wick rotation, under which Euclidean form factors generally acquire an imaginary part due to logarithmic contributions for $q_M^2>0$. The asymptotic analysis shows that the real part of the Minkowskian form factors coincides with the Euclidean one up to the sign of $w_M$. Away from the asymptotic regimes, the continuation must be performed numerically by fitting the Euclidean data, a procedure that may introduce spurious structures such as artificial poles or zeros. However, for all admissible interpolating functions the continuation is stable under changes of the fitting ansatz, indicating that the Minkowskian form factors are robust and that the intermediate-momentum region is under numerical control within the present reconstruction scheme.

The Minkowskian background propagators exhibit a single pole at $q^2=0$ and acquire both real and imaginary parts. The real part remains positive over the entire momentum range, while $\mathrm{Im}[G_{\mathrm{TT}}]$ is positive only in the IR, in agreement with previous studies \cite{Pawlowski:2025etp,Bonanno:2021squ,Fehre:2021eob,Pastor-Gutierrez:2024sbt,Kher:2025rve,article}. This behavior does not signal a violation of unitarity. In the presence of non-local form factors, a Källén--Lehmann representation with a unique spectral density is generally not guaranteed; rather, unitarity is controlled by the massless pole of $1/(q^2+i\epsilon)$, yielding the standard spectral density $\rho(\mu^2)=\delta(\mu^2)/\pi$. Different choices of $c_{\mathrm{R}}$ and $c_{\mathrm{Ricc}}$ do not modify these conclusions, but only affect the UV asymptotic behavior of the propagators, as in the Euclidean case.

After analytical continuation at $k=0$, the effective Newtonian coupling becomes complex-valued. Its real part closely follows the Euclidean behavior, while the imaginary part reflects the discontinuity of the propagator across the branch cut generated by the logarithmic terms.

The corrections to the Newtonian potential are obtained from the Fourier transform of the propagators. Reference AS propagators, which contain additional poles, lead to oscillatory potentials that are difficult to interpret physically. By contrast, non-reference AS propagators yield well-behaved corrections in which the $1/r$ singularity is softened and the potential approaches a finite limit as $r\to0$. This may indicate a resolution of the singularity beyond the weak-field approximation, although a definitive conclusion would require a fully non-linear analysis.

Overall, although obtained at the background level, the results presented here support the asymptotic safety scenario and are consistent also with the expectation that non-local dynamics may address the ghost problem \cite{Platania:2020knd,Becker:2017tcx,Draper:2020bop,Draper:2020knh} and soften the singularity at $r=0$ \cite{Basile:2024oms,Bosma:2019aiu,Knorr:2021iwv}. Nevertheless, further work is required before definitive conclusions can be drawn \cite{Basile:2024oms}.

The asymptotically safe form factors derived in \cite{Glaviano:2026lew} are obtained within a background approximation. Consequently, the results presented here capture the leading non-local quantum effects but neglect backreaction effects and the full dynamics of spacetime fluctuations. Extending the present analysis to fluctuation correlators and more general truncations therefore remains an important direction for future work.

\section*{Acknowledgements}
\noindent Emiliano would like to thank Alfio Bonanno for his valuable support and encouragement throughout this work.

\newpage

\begin{appendices}

\section{Hessian of the non-local theory}\label{appderbeta}
\noindent In this appendix we sum up the technical detail to derive the Hessian.

We adopt the background field method where we decompose the metric using the linear split
\begin{equation}
g_{\mu\nu}=\bar g_{\mu\nu}+h_{\mu\nu}\,,
\end{equation}
where barred quantities denote background fields.

For convenience, we introduce the parametrization
\begin{equation}
\frac{1}{16\pi G_k}=\frac{Z_{Nk}}{16\pi G_0}=2\kappa^2 Z_{Nk}\,,
\end{equation}
where $Z_{Nk}$ denotes the wave-function renormalization associated with Newton's constant.

The gravitational action is supplemented by the standard gauge-fixing term:
\begin{equation}
S_{gf}\left[h;\bar{g}\right]=\frac{1}{\alpha}\int{d^dx\sqrt{\bar{g}}{\bar{g}}^{\mu\nu}F_\mu\left(h;\bar{g}\right)F_\nu\left(h;\bar{g}\right)}
\end{equation}
where
\begin{equation}\begin{split}
&F_\mu\left(h;\bar{g}\right)=\kappa\mathcal{F}_\mu^{\alpha\beta}\left[\bar{g}\right]h_{\alpha\beta},\\ &\mathcal{F}_\mu^{\alpha\beta}\left[\bar{g}\right]h_{\alpha\beta}=\left(\delta_\mu^\beta{\bar{g}}^{\alpha\gamma}{\bar{\nabla}}_\gamma-\bar{\omega}{\bar{g}}^{\alpha\beta}{\bar{\nabla}}_\mu\right)h_{\alpha\beta}={\bar{\nabla}}^\alpha h_{\alpha\mu}-\bar{\omega}{\bar{\nabla}}_\mu h
\end{split}\end{equation}
here $\alpha$ and $\bar\omega$ are the gauge parameters.

The gauge fixing yields the standard ghost action term
\begin{equation}
S_{gh}\left[h;\bar{g}\right]=-\frac{1}{\kappa}\int{d^dx\sqrt{\bar{g}}{\bar{C}}_\mu{\bar{g}}^{\mu\nu}\frac{\partial F_\nu}{\partial h_{\alpha\beta}}\mathcal{L}_C\left({\bar{g}}_{\alpha\beta}+h_{\alpha\beta}\right)}
\end{equation}
the Lie derivative is given by $\mathcal{L}_vT_{\alpha\beta}=v^\rho\partial_\rho T_{\alpha\beta}+T_{\rho\beta}\partial_\alpha v^\rho+g_{\rho\alpha}\partial_\beta v^\rho$. 

The Hessian of Eq.~(\ref{TE}) is obtained by expanding the action to second order in the metric fluctuations. The resulting expression reads \cite{Glaviano:2026lew}
\begin{equation}
\label{secvaract}
\Gamma_k^{\left(2\right)}\left[g\right]=\Gamma_k^{\left(2\right),EH}\left[g\right]+\Gamma_{k,f}^{\left(2\right)}\left[g\right]+\Gamma_{k,\delta f}^{\left(2\right)}\left[g\right]+S_{gh}^{\left(2\right)}
\end{equation}
where the first term corresponds to the Einstein--Hilbert contribution, the second and third terms encode the form-factor contributions, and the last term is the ghost Hessian.

The Einstein-Hilbert term is given by
\begin{equation}\begin{split}
&\Gamma_{k}^{(2),EH}\left[g\right]=\kappa^2Z_{Nk}\int{d^dx}\sqrt{\bar{g}}h_{\mu\nu}\Bigg\{-K_{\rho\sigma}^{\mu\nu}\left(\alpha;\omega\right)\Box+U_{\rho\sigma}^{\mu\nu}\left(-\Box\right)+\\
&+\left(1-\frac{2f_k^{\left(\mathrm{R}\right)}\left(-\Box\right)R}{\kappa^2Z_{Nk}}-\frac{1}{\alpha}\right)\delta_\rho^\nu{\bar{\nabla}}^\mu{\bar{\nabla}}_\sigma-\left(1-\frac{2f_k^{\left(\mathrm{R}\right)}\left(-\Box\right)R}{\kappa^2Z_{Nk}}-\frac{2\bar{\omega}}{\alpha}\right){\bar{g}}^{\mu\nu}\nabla_\rho{\bar{\nabla}}_\sigma\Bigg\}h^{\rho\sigma}
\end{split}\end{equation}
where
\begin{equation}\begin{split}
&K_{\rho\sigma}^{\mu\nu}\left(\alpha,\bar{\omega}\right)=\frac{1}{4}\left(1-\frac{2f_k^{\left(\mathrm{R}\right)}\left(-\Box\right)R}{\kappa^2Z_{Nk}}\right)\left(\delta_\rho^\mu\delta_\sigma^\nu+\delta_\sigma^\mu\delta_\rho^\nu\right)-\left(\frac{1}{2}\left(1-\frac{2f_k^{\left(\mathrm{R}\right)}\left(-\Box\right)R}{\kappa^2Z_{Nk}}\right)-\frac{{\bar{\omega}}^2}{\alpha}\right){\bar{g}}^{\mu\nu}{\bar{g}}_{\rho\sigma}\\
&U_{\rho\sigma}^{\mu\nu}\left(-\Box\right)=\frac{1}{4}\left(\delta_\rho^\mu\delta_\sigma^\nu+\delta_\sigma^\mu\delta_\rho^\nu-g^{\mu\nu}g_{\rho\sigma}\right)\left(R-2{\bar{\lambda}}_k-\frac{Rf_k^{\left(\mathrm{R}\right)}\left(-\Box\right)R}{\kappa^2Z_{Nk}}-\frac{R_{ab}f_k^{\left(\mathrm{Ricc}\right)}\left(-\Box\right)R^{ab}}{\kappa^2Z_{Nk}}\right)+\\
&+\left[-\frac{1}{4}\left(R_\sigma^\nu\delta_\rho^\mu+R_\sigma^\mu\delta_\rho^\nu+R_\rho^\nu\delta_\sigma^\mu+R_\rho^\mu\delta_\sigma^\nu\right)+g^{\mu\nu}R_{\rho\sigma}-\frac{1}{2}\left(R_{\rho\sigma}^{\mu\nu}+R_{\sigma\rho}^{\mu\nu}\right)\right]\left(1-\frac{2f_k^{\left(\mathrm{R}\right)}\left(-\Box\right)R}{\kappa^2Z_{Nk}}\right)
\end{split}\end{equation}

The second term can be written in the form
\begin{equation}
\Gamma_{k,f}^{\left(2\right)}\left[g\right]=\int{d^dx\sqrt g\left\{h_{\mu\nu}\left[H_{\rho\sigma}^{\mu\nu}\left(-\Box\right)+\left(f_k^{\left(\mathrm{Ricc}\right)}\left(-\Box\right)R^{ab}\right)L_{\rho\sigma ab}^{\mu\nu}\left(-\Box\right)\right]h^{\rho\sigma}\right\}}
\end{equation}
where
\begin{equation}\begin{split}
&H_{\rho\sigma}^{\mu\nu}\left(-\bar\Box\right)=2\left(\bar R^{\mu\nu}-\bar\nabla^\mu\bar\nabla^\nu+\bar g^{\mu\nu}\bar\Box\right)f_k^{\left(\mathrm{R}\right)}\left(-\bar\Box\right)\left(\bar R_{\rho\sigma}-\bar\nabla_\rho\bar\nabla_\sigma+\bar g_{\rho\sigma}\bar\Box\right)+\\
&+\frac{1}{2}\left(\bar g^{\mu\nu}\bar\nabla_a\bar\nabla_b-\bar g_b^\mu\bar\nabla_a\bar\nabla^\nu-\bar g^\mu_{\ a}\bar\nabla_b\bar\nabla^\nu+\bar g_a^\nu\bar g_b^\mu\bar\Box\right)f_k^{\left(\mathrm{Ricc}\right)}\left(-\bar\Box\right)\left(\bar g_{\rho\sigma}\bar\nabla^a\bar\nabla^b-\bar g_\sigma^b\bar\nabla^a\bar\nabla_\rho-\bar g_\sigma^a\bar\nabla^b\bar\nabla_\rho+\bar g_\rho^a\bar g_\sigma^b\bar\Box\right)\\
&L_{\rho\sigma ab}^{\mu\nu}\left(-\bar\Box\right)=\bar g^{\mu\nu}
\left(-\bar g_{\rho\sigma}\bar\nabla_a\bar\nabla_b+\bar g_{\sigma b}\bar\nabla_a\bar\nabla_\rho+\bar g_{\sigma a}\bar\nabla_b\bar\nabla_\rho-\bar g_{\rho a}\bar g_{\sigma b}\bar\Box\right)+\bar g_{a\sigma}\left(\bar g^{\mu\nu}\bar g_{b\rho}+\bar g_b^\nu\bar g_\rho^\mu\right)\bar\Box\\
&+\bar g_\rho^\mu\bar g_\sigma^\nu\bar\nabla_a\bar\nabla_b-\bar g_a^\nu\bar g_{\rho\sigma}\bar\nabla_b\bar\nabla^\mu-\bar g_b^\nu\bar g_{\rho\sigma}\bar\nabla_a\bar\nabla^\mu\\
&-2\Big(\bar g_{b\rho}\bar g_\sigma^\mu\bar\nabla^\nu\bar\nabla_a+\bar g_a^\mu\big(\bar g_b^\nu\bar\nabla_\sigma\bar\nabla_\rho+\bar g_\rho^\nu\bar\nabla_b\bar\nabla_\sigma\big)+\bar g_{a\rho}\big(\bar g_\sigma^\mu\bar\nabla^\nu\bar\nabla_b-\bar g_{b\sigma}\bar\nabla^\nu\bar\nabla^\mu\big)+\bar g_b^\nu\big(\bar g_\rho^\mu\bar\nabla_a\bar\nabla_\sigma+\bar g_{a\rho}\bar\nabla_\sigma\bar\nabla^\mu\big)\Big)\,.
\end{split}\end{equation}
In flat spacetime, $f_k^{(Ric)}(-\Box) R^{ab}=0$, and the derivatives in $H_{\rho\sigma}^{\mu\nu}(-\Box)$ commute with the form factors. This reproduces the Hessian presented in  \cite{Biswas:2011ar,Frolov:2015usa,Biswas:2013cha,Modesto:2011kw}.


The third piece is the term involving the variation of the form factors. Calling $\mathcal{R}=\{R,R_{\mu\nu}\}$ this piece is given schematically by:
\begin{equation}
\Gamma_{k,\delta f}^{\left(2\right)}\left[g\right]=\int d^dx\sqrt g\,\mathcal{R}_1\,\delta f\left(\Delta\right)\mathcal{R}_2+\int d^dx\sqrt g\,\mathcal{R}_1\,\delta^2 f\left(\Delta\right)\mathcal{R}_2
\end{equation}
The two contributions are given, respectively, by
\begin{equation}\begin{split}
&\int d^dx\sqrt g\,\mathcal{R}_1\,\delta f\left(\Delta\right)\mathcal{R}_2
=\int d^dx\sqrt g\,\frac{f\left(\Delta_1\right)-f\left(\Delta_2\right)}{\Delta_1-\Delta_2}\left(\delta\Delta_2\right)\mathcal{R}_1\mathcal{R}_2,\\
&\int d^dx\sqrt g\,\mathcal{R}_1\,\delta^2 f\left(\Delta\right)\mathcal{R}_2=\\
&\quad=\int d^dx\sqrt g\,\bigg\{
2\left[\frac{f\left(\Delta_1\right)-f\left(\Delta_2\right)}{\left(\Delta_1-\Delta_2\right)^2}
-\frac{f'\left(\Delta_2\right)}{\Delta_1-\Delta_2}\right]\left(\delta\Delta_2\right)^2+\frac{f\left(\Delta_1\right)-f\left(\Delta_2\right)}{\Delta_1-\Delta_2}\,\delta^2\Delta_2
\bigg\}\mathcal{R}_1\mathcal{R}_2.
\end{split}\end{equation}
Here a prime denotes a derivative with respect to $\Delta=-\Box$, and the subscripts $1$ and $2$ indicate on which curvature tensor the operator $\Box$ acts. The explicit expressions in terms of $h_{\mu\nu}$ are rather lengthy and will not be displayed here.

The Hessian of ghost action is given by
\begin{equation}
S^{(2)}_{gh}=-\int{d^dx\sqrt{\bar{g}}\left[{\bar{C}}_\mu\left(-\bar \Box+\frac{\bar{R}}{d}\right)C^\mu+{\bar{C}}_\mu\left(1-2\bar{\omega}\right){\bar{\nabla}}^\mu{\bar{\nabla}}_\nu C^\nu\right]}
\end{equation}
In this work we choose the Lorentz gauge $\alpha=1$ and $\bar{\omega}=1/2$.

\section{Techniques to derive the asymptotics of the Newtonian potential}\label{appNewpot}
\noindent In this appendix we outline how to derive the asymptotic corrections to the Newtonian potential without computing the full Fourier transform. 

The first step is to isolate the pole structure of the propagator $G(p^2)$ via the decomposition
\begin{equation}
\label{spectraldec}
G\left(q^2\right)=\sum_{n=1}^{N}\frac{\alpha_n}{q^2-q_n^2}+h\left(q^2\right),
\end{equation}
where $\alpha_n$ are the residues at the poles $q^2=q_n^2$, and $h(q^2)$ denotes the regular (analytic) part.\footnote{This decomposition should not be confused with a purely algebraic one, where similar expressions may arise but the poles do not necessarily correspond to physical states.} The Fourier transform of the pole contributions yields the standard Coulomb, Yukawa, or oscillatory potentials, depending on the sign of $q_n^2$, whereas the non-trivial asymptotic behavior is entirely encoded in $h(q^2)$.

The Newtonian potential is obtained from
\begin{equation}
\label{potnew}
V\left(r\right)=-\frac{m}{4}\int{\frac{d^3p}{\left(2\pi\right)^3}e^{-ip\cdot r}G_{0000}\left(p^2\right)}=-\frac{m}{8\pi^2}\int{dp\,\frac{\sin{\left(pr\right)}}{pr}\,p^2\,G_{0000}\left(p^2\right)},
\end{equation}
where $G_{0000}(p^2)$ is given in eq.~(\ref{G0000}). In the following we restrict the analysis to shifted asymptotically safe form factors, since the reference solutions lead to Euclidean propagators with real positive poles and consequently to oscillatory potentials without a clear physical interpretation.

The isolation of the pole contributions is essential for the analysis. If the poles were not separated explicitly, the integrand would diverge at their locations, and extracting the asymptotic behaviour without performing the full integral would generally lead to spurious results.

To extract the asymptotics, it is convenient to split the integral into infrared and ultraviolet contributions,
\begin{equation}
V\left(r\right)=\int_{0}^{\Lambda_p}{\frac{\sin{\left(pr\right)}}{pr}M_{fi}\left(p\right)\,dp}+\int_{\Lambda_p}^{\infty}{\frac{\sin{\left(pr\right)}}{pr}M_{fi}\left(p\right)\,dp}
\equiv V_{IR}\left(r,\Lambda_p\right)+V_{UV}\left(r,\Lambda_p\right),
\end{equation}
where $M_{fi}(p)$ is the non-relativistic scattering amplitude and $\Lambda_p$ is a separation scale defined by
\begin{equation}
\Lambda_p\, r \sim 1.
\end{equation}
In the large-$r$ regime the dominant contribution arises from $V_{IR}$, whereas for small $r$ it is governed by $V_{UV}$, as shown below.

\subsection{The large \texorpdfstring{$r$}{N} regime}
In the large-$r$ regime, the condition $\Lambda_p r\sim1$ implies that the dominant contribution arises from the low-momentum region $p\lesssim1/r$. By the Riemann--Lebesgue theorem, the potential vanishes for $r\to\infty$. Accordingly, taking $\Lambda_p\to\infty$ selects the full IR contribution, while the UV part vanishes due to the regularity of the integrand at large momenta, i.e. $V_{UV}(r,\Lambda_p \to \infty)\to 0$ so
\begin{equation}
V_{IR}\left(r\right)\equiv\lim_{\Lambda_p\rightarrow\infty}V\left(r\right)=\lim_{\Lambda_p\rightarrow\infty}V_{IR}\left(r,\Lambda_p\right).
\end{equation}
Substituting the small-momentum expansion of the propagator, eq.~(\ref{propcrcriccsmallp}), into the IR contribution, we obtain
\begin{equation}\begin{split}
&V_{IR}\left(r\right)=\\
&=-\frac{m}{8\pi^2}\frac{1}{r}\int_{0}^{\infty}\left[\frac{\sin\left(pr\right)}{\frac{p}{16\pi G}}+\frac{1}{m_p^3}\sum_{n=1}^{+\infty}\left(a_n+\sum_{m=1}^{n}2^m b_{nm}\ln^m \left(\frac{p}{m_p}\right)\right)\left(\frac{p}{m_p}\right)^{2n-1}\sin\left(\left(\frac{p}{m_p}\right)m_pr\right)\right]dp.
\end{split}\end{equation}
Although these integrals are formally divergent in the classical sense, they are well-defined as distributions:
\begin{equation}\begin{split}
\label{distrint}
&\int_{0}^{\infty}p^{2n}\sin\left(pr\right)\,dp=\frac{(-1)^n(2n)!}{r^{2n+1}},\\
&\int_{0}^{\infty}p^{2n+1}\sin\left(pr\right)\,dp=\frac{\left(-1\right)^n\left(2n+1\right)!\delta^{\left(2n+1\right)}\left(r\right)}{r^{2\left(n+1\right)}},\\
&\int_{0}^{\infty}p^n\ln^m(p)\sin\left(pr\right)\,dp=\frac{\partial^m}{\partial n^m}\int_{0}^{\infty}p^n\sin\left(pr\right)\,dp.
\end{split}\end{equation}
The last class of integrals can be written explicitly as
\begin{equation}\begin{split}
&\int_{0}^{\infty}p^n\ln^m(p)\sin\left(pr\right)\,dp=\\
&=2^{m}\sum_{l=0}^{m}\sum_{q=0}^{m}\,r^{-n-1}\frac{(-1)^l m!}{l!(m-l-q)!q!}\left(\frac{\pi}{2}\right)^q \frac{1+(-1)^{n+q}}{2}(-1)^{\frac{n+q}{2}}\ln^l(r)\left.\frac{\partial^{m-l-q}\Gamma(x+1)}{\partial x^{m-l-q}}\right|_{x=n}.
\end{split}\end{equation}
The odd polynomial integrals generate contact terms proportional to derivatives of the Dirac delta function, which vanish for $r\neq0$. Therefore, only the even contributions determine the large-distance behaviour. Using these results, after straightforward algebra, we obtain
\begin{equation}
\label{newpotIR}
V_{IR}\left(r\right)=-\frac{mG}{r}\left[1+\frac{1}{8\pi^2}\sum_{n=1}^{+\infty}\left(B_{n0}+\sum_{m=1}^{n-1}B_{nm}\ln^m\left(m_pr\right)\right)\frac{1}{(m_pr)^{2n}}\right],
\end{equation}
where
\begin{equation}\begin{split}
B_{nm}=\sum_{l=m}^{n}2^{l}b_{nl}\sum_{q=0}^{m}\frac{(-1)^l m!}{l!(m-l-q)!q!}\left(\frac{\pi}{2}\right)^q
\frac{(-1)^q-1}{2}(-1)^{\frac{1+2n+q}{2}}
\left.\frac{d^{m-l-q}\Gamma(z+1)}{dz^{m-l-q}}\right|_{z=2n-1}.
\end{split}\end{equation}
evaluating the expression explicitly reproduces eq.(\ref{newpotIRexp}).

\subsection{The small \texorpdfstring{$r$}{N} regime}
In the small-$r$ domain, the condition $\Lambda_p r \sim 1$ implies that the dominant contribution arises from the large-momentum region, $p \gg 1/r$. Sending $\Lambda_p \to 0$ isolates the full UV contribution. Since the integrand is regular at $p=0$, the IR part vanishes, ${V}_{IR}(r,\Lambda_p \to 0) \to 0$, and therefore
\begin{equation}
V_{UV}(r \to 0) \equiv \lim_{\Lambda_p \to 0} V(r) = \lim_{\Lambda_p \to 0} V_{UV}(r,\Lambda_p)\,.
\end{equation}
In contrast to the large-$r$ case, the Dirac delta contributions cannot be neglected and yield a well-defined contact term at $r=0$. Their total contribution is equivalent to evaluating eq.~(\ref{potnew}) at $r=0$, while excluding the delta terms from $V_{UV}(r\neq 0)$. The contact term is given by
\begin{equation}
\label{c0newtpot}
V_{UV}(r=0) = -\frac{m}{8\pi^2} \int_{0}^{+\infty} p^2 G_{0000}(p^2)\, dp\,,
\end{equation}
and the full UV expansion reads
\begin{equation}
V_{UV}^{\text{tot}}(r \to 0) = V_{UV}(r=0) + V_{UV}(r \neq 0)\,.
\end{equation}
The term at $r=0$ cannot be computed analytically; however, dimensional analysis implies
\begin{equation}
V_{UV}(r=0) = -\frac{m}{8\pi^2 m_p}\, c\,,
\end{equation}
where $c$ is a dimensionless constant that can be determined numerically. For the contribution at $r \neq 0$, inserting the large-momentum expansion in eq.~(\ref{propcrcricclargep}) into the UV integral yields
\begin{equation}
V_{UV}(r) = -\frac{m}{8\pi^2 r} \int_{0}^{\infty} \sin\!\left(\left(\frac{p}{m_p}\right) m_p r\right)
\left[
\frac{1}{m_p^2} \sum_{n=0}^{+\infty}
\left(a_n + \sum_{m=1}^{n} 2^m a_{nm} \ln^m\left(\frac{p}{m_p}\right)\right)
\left(\frac{m_p}{p}\right)^{2n+3}
\right] dp\,.
\end{equation}
The distributional identities in eq.~(\ref{distrint}) can be extended to negative values of $n$:
\begin{equation}\begin{split}
&\int_{0}^{\infty} \frac{\sin(pr)}{p^{2n+1}}\, dp = \frac{(-1)^n \pi r^{2n}}{2 \Gamma(1+2n)}\,,\\
&\int_{0}^{\infty} p^{-n} \ln^m(p)\, \sin(pr)\, dp = \\
&=\frac{2^m}{(-1)^m} \sum_{l=0}^{m} \sum_{q=0}^{m}r^{n-1}\frac{(-1)^l m!}{l!\,(m-l-q)!\,q!}\left(\frac{\pi}{2}\right)^q\frac{1+(-1)^{q-n}}{2}(-1)^{\frac{q-n}{2}}\ln^l(r)\,\left.\frac{\partial^{m-l-q} \Gamma(x+1)}{\partial x^{m-l-q}}\right|_{x=-n}\,.
\end{split}\end{equation}
Using these expressions, one obtains
\begin{equation}
\label{newpotUV}
V_{UV}(r) = -\frac{m}{8\pi^2 m_p}\left\{c + (m_p r)\left[-\frac{a_0 \pi}{8}+ \sum_{n=1}^{+\infty}\left(a_n \frac{(-1)^{n+1} \pi}{2(2(n+1))!}+ \sum_{m=0}^{n} B_{nm} \ln^m(m_p r)\right)(m_p r)^{2n}\right]\right\}\,,
\end{equation}
where
\begin{equation}\begin{split}
B_{nm} = \sum_{l=1}^{n-m} a_{nl}\frac{2^{-m}}{(-1)^m}\sum_{q=0}^{m}\frac{(-1)^l m!}{l!\,(m-l-q)!\,q!}\left(\frac{\pi}{2}\right)^q\frac{1+(-1)^{q-(2n+3)}}{2}(-1)^{\frac{q-(2n+3)}{2}}\left.\frac{\partial^{m-l-q} \Gamma(x+1)}{\partial x^{m-l-q}}\right|_{x=2n+3}\,.
\end{split}\end{equation}
The UV series is finite at $r=0$. Evaluating explicitly this expression reproduces eq. (\ref{newpotUVexp}). 

Setting $r=1/q$, the limit $r\to0$ corresponds to a large-momentum expansion in $q$. In this regime, expanding before performing the exact $p$ integral removes the contribution from the region $q^2\le p^2$, requiring a regularization procedure to reconstruct the full result. This is analogous to the situation discussed in \cite{Glaviano:2026lew} for the large-$q$ behavior of the form factors. Applying the same technique yields Eq.~(\ref{newpotUV}) without explicitly using distributional integrals, with the divergent contributions canceling as in \cite{Glaviano:2026lew}.

\section{The flow of the form factors}\label{App:asymFF}
In this Appendix we collect some results of \cite{Glaviano:2026lew} used in this work.

\subsection{The running couplings and the form factor}
The proper-time flow equation used in \cite{Glaviano:2026lew} is given by \cite{Bonanno:2019ukb}
\begin{equation}
\label{PTFE}
k\partial_k \Gamma_k [\phi] = \frac{1}{2}\int_{0}^{\infty}\frac{ds}{s}\rho\left(s,k^2 Z_k\right)\mathrm{STr}\left[e^{-s\Gamma_k^{(2)}}\right],
\end{equation}
where $k$ denotes the coarse graining scale, $\Gamma_k^{(2)}$ is the Hessian of the action, $Z_k$ the wavefunction renormalization, and $\rho(s,k^2Z_k)$ a cutoff function. This is given by \cite{Bonanno:2019ukb}
\begin{equation}
\label{cutoff}
\rho_m(s,k^2 Z_k) = \left(2 + \epsilon\frac{k \partial_k Z_k}{Z_k}\right)\frac{(s\, m\, k^2 Z_k)^m}{\Gamma(m)}e^{-s\, m\, k^2 Z_k}\,,
\end{equation}
where $m>0$ controls the behavior of $\rho(s,k^2 Z_k)$ in the interpolating region. The parameter $\epsilon$ distinguishes between two types of scheme: type B scheme ($\epsilon=1$) and type C scheme ($\epsilon=0$). 

The choice $m=d/2+1$ gives the same flow as the Wetterich equation with an optimized regulator. In $d=4$ one finds $m=3$, which is the value used in this work \cite{Glaviano:2026lew}.

The flow equations for the Newton coupling and the form factors in $d=4$ with $m=3$ read
\begin{equation}\begin{split}
&k\partial_k g_k=g_k\left(2+\eta_k\right),\quad\quad \eta_k=-\frac{46g_k}{4\pi-13\epsilon g_k}\\
&k\partial_kF_k^{(i)}\left(x\right)=2x\partial _xF_k^{(i)}\left(x\right)+H^{(i)}\left(\eta_k,x\right)
\end{split}\end{equation}
where $i=\mathrm{R}$, $\mathrm{Ricc}$, $x=q^2/k^2$ and
\begin{equation}\begin{split}
&H^{\left(\mathrm{R}\right)}\left(\eta_k,x\right)=-\frac{x^3+78x^2-72x+1296}{32\pi^2x^2\left(x+12\right)^2}+\frac{\left(25x^3+438x^2+2412x+3240\right)\left(2-\epsilon\eta_k\right)}{128\pi^2x^2\left(x+12\right)^2}-\\
&-\left(\frac{27\left(13x^2+144x+180\right)\left(2-\epsilon\eta_k\right)}{16\pi^2x^\frac{5}{2}\left(x+12\right)^\frac{5}{2}}+\frac{27\left(5x^2-72\right)}{4\pi^2x^\frac{5}{2}\left(x+12\right)^\frac{5}{2}}\right)\tanh^{-1}\left(\frac{\sqrt x}{\sqrt{x+12}}\right)\\
&H^{\left(\mathrm{Ricc}\right)}\left(\eta_k,x\right)=\frac{x^3+6x^2+1008x+5184}{16\pi^2x^2\left(x+12\right)^2}-\frac{\left(13x^3+78x^2+4032x+12960\right)(2-\epsilon\eta_k)}{64\pi^2x^2\left(x+12\right)^2}+\\
&+\left(\frac{27\left(35x^2+264x+720\right)\left(2-\epsilon\eta_k\right)}{8\pi^2x^\frac{5}{2}\left(x+12\right)^\frac{5}{2}}-\frac{27\left(7x^2+72x+288\right)}{2\pi^2x^\frac{5}{2}\left(x+12\right)^\frac{5}{2}}\right)\tanh^{-1}\left(\frac{\sqrt x}{\sqrt{x+12}}\right)
\end{split}\end{equation}
In the mixed scheme (MS), the Newton coupling is evaluated using a type-C cutoff ($\epsilon=0$), while the form factors are computed with a type-B cutoff ($\epsilon=1$). In the full B-scheme, both sectors are treated with a type-B cutoff. 

In the mixed scheme, the running couplings $g_k$ and its correspondent anomalous dimension $\eta_k$ are given by 
\begin{equation}
\label{unimproGeta}
g_k=\frac{4k^2\pi}{23k^2+4m_p^2\pi},\quad \quad \eta_k=-\frac{46k^2}{23k^2+4m_p^2\pi}
\end{equation}
where $m_p$ is the Plank mass.


The integrated solutions of the form factor equation are controlled by an IR Gaussian regime for $k\ll m_p$ and UV non-Gaussian fixed point for $k\gg m_p$ but generically exhibit a logarithmic dependence on the UV scale $\Lambda$. This dependence is eliminated by imposing a renormalization condition that selects trajectories whose UV behavior is governed by the non-Gaussian fixed point. The resulting solution reads:
\begin{equation}\begin{split}
\label{ASsoladim}
&F_k\left(x;\Lambda,x_{\mathrm{BC}}\right)=\bar F_\ast^{\mathrm{NGFP}}\left(x;x_{\mathrm{BC}}\right)+\delta F_k(x,\Lambda)\\
&\delta F_k(x,\Lambda)\equiv -\int_{\frac{k^2x}{\Lambda^2}}^{x}\left[\frac{H\left(\eta\left(k\sqrt{\frac{x}{v}}\right),v\right)}{2v}-\frac{H\left(\eta_\ast=-2,v\right)}{2v}\right]dv
\end{split}\end{equation}
which has a smooth $\Lambda \to \infty$ limit and preserves the AS properties of the flow. Here
\begin{equation}
\bar F_\ast^{\mathrm{NGFP}}\left(x;x_{\mathrm{BC}}\right)=c_0-\int_{x_{\mathrm{BC}}}^{x}\frac{H\left(\eta_\ast=-2,v\right)}{2v}\equiv c_0+F_\ast^{\mathrm{NGFP}}\left(x;x_{\mathrm{BC}}\right)
\end{equation}
is the solution of the fixed-point equation, with integration constant $c_0$ fixed by the boundary condition $F_\ast^{\mathrm{NGFP}}(x_{\mathrm{BC}})=c_0$. The shift $c_0$ is fixed by specifying the value of the form factor at infinite momentum, $x_{BC}\to\infty$ \cite{Knorr:2021niv}.

Comparing with eq.~(\ref{ASflow}), the reference solution and boundary function are identified as
\begin{equation}\begin{split}
\label{flowcomp}
&F_{k}^{(\mathrm{AS})}(x;x_{\mathrm{BC}})=F_\ast^{\mathrm{NGFP}}(x;x_{\mathrm{BC}}) + \lim_{\Lambda\to\infty}\delta F_k(x,\Lambda),\\
&h(x_{\mathrm{BC}})= c_0.
\end{split}\end{equation}
Fig.~\ref{plotflowad} shows the flow of $F_{k}^{(\mathrm{AS})}(x;x_{\mathrm{BC}})$ for $c_0=0$, while Fig.~\ref{plotFFASeps1} shows the corresponding dimensional solution at $k=0$. These solutions correspond to those used in Sec.~\ref{secEFFgamma2} and Sec.~\ref{seceucprop}, respectively. The analytically continued version is instead used in Sec.~\ref{secAnalcont}.

\subsection{The asymptotic expansions in \texorpdfstring{$x$}{N} for the asymptotically safe form factors}
The asymptotically safe form factors can be analyzed analytically in both the small- and large-$x$ regimes. In the small-$x$ regime, they read
\begin{equation}
\label{aeFASsmallk}
F_k\left(x\ll1\right)=a_0\left(x_{\mathrm{BC}}\right)+a\ln{(x)}+c_0\ln{\left(1+\frac{4m_p^2\pi}{23k^2}\right)}+\sum_{i=1}^{+\infty}{\left[\sum_{j=0}^{i-1}{b_{ij}\left(\frac{k^2}{m_p^2}\right)^j}+c_i\left(\frac{k^2}{m_p^2}\right)^i\ln{\left(1+\frac{4m_p^2\pi}{23k^2}\right)}\right]x^i}
\end{equation}
in Sec. \ref{secEFFgamma2} the expansions for $F_k^{(\mathrm{Ricc})}(x)$ are used. The first terms of this expansion are given by
\begin{equation}\begin{split}
&F_k\left(x\ll1\right)=c_{\mathrm{Ricc}}+\frac{7}{120}\ln{\left(x_{\mathrm{BC}}\right)}+\frac{29160+x_{\mathrm{BC}}\left(13500+\left(1158-121x_{\mathrm{BC}}\right)x_{\mathrm{BC}}\right)}{3600\pi^2x_{\mathrm{BC}}^2\left(12+x_{\mathrm{BC}}\right)}+\\
&+\frac{-5832-7x_{\mathrm{BC}}\left(432+x_{\mathrm{BC}}\left(54+x_{\mathrm{BC}}\left(18+x_{\mathrm{BC}}\right)\right)\right)}{60\pi^2x_{\mathrm{BC}}^{\frac{5}{2}}\left(12+x_{\mathrm{BC}}\right)^\frac{3}{2}}\mathrm{tanh}^{-1}\left(\sqrt{\frac{x_{\mathrm{BC}}}{12+x_{\mathrm{BC}}}}\right)+\\
&+\frac{7}{192\pi^2}\ln{\left(1+\frac{4m_p^2\pi}{23k^2}\right)}-\frac{7\ln{\left(x\right)}}{120\pi^2}+x\left(\frac{17}{2688\pi^2}+\frac{3013k^2}{53760m_p^2\pi^3}\ln{\left(1+\frac{4m_p^2\pi}{23k^2}\right)}\right)+\\
&+x^2\left(-\frac{181}{362880\pi^2}-\frac{1265k^2}{145152m_p^2\pi^3}+\frac{29095k^4}{580608m_p^4\pi^4}\ln{\left(1+\frac{4m_p^2\pi}{23k^2}\right)}\right)+O(x^3)
\end{split}\end{equation}
in the limits $k/m_p\ll1$ and $k\gg1$ the expansion correctly includes the fixed-point regimes in the small $x$ domain. This expansion is used in eq. (\ref{IRefateff}) and eq. (\ref{serieIRgeff}).

In the large $x$ regime, the solution is given by
\begin{equation}\begin{split}
\label{aeFASlargek}
&F_k\left(x\gg1\right)=a_0(x_{\mathrm{BC}})+\left[-\frac{3}{8\pi^2}+\frac{m_p^2}{k^2}\left(\frac{137}{1104\pi}+\frac{13\ln{\left[\frac{3}{x}\left(1+\frac{4\pi m_p^2}{23k^2}\right)\right]}}{368\pi}\right)\right]\frac{1}{x}+\\
&+\left[\frac{27}{8\pi^2}-\frac{117m_p^2}{184\pi k^2}+\frac{m_p^4}{k^4}\left(\frac{9}{529}+\frac{117\ln{\left[\frac{3}{x}\left(1+\frac{4\pi m_p^2}{23k^2}\right)\right]}}{1058}\right)\right]\frac{1}{x^2}+\\
&\bigg[-\frac{57}{\pi^2}+\frac{63\ln{\left(\frac{x}{3}\right)}}{2\pi^2}+\frac{m_p^2}{k^2}\left(\frac{2943}{368\pi}-\frac{945\ln{\left(\frac{x}{3}\right)}}{184\pi}\right)+\frac{m_p^4}{k^4}\left(-\frac{999}{529}+\frac{945\ln{\left(\frac{x}{3}\right)}}{529}\right)+\\
&\resizebox{1.05\hsize}{!}{$+\frac{m_p^6}{k^6}\left(\frac{7776\pi\ln{\left[\frac{3}{x}\left(1+\frac{4\pi m_p^2}{23k^2}\right)\right]}}{12167}-\frac{108\pi\left(35{\rm \operatorname{Li}}_2\left(-\frac{4\pi m_p^2}{23k^2}\right)-24\right)}{12167}-\frac{3780\pi\log{\left(\frac{x}{3}\right)}\ln{\left(1+\frac{4\pi m_p^2}{23k^2}\right)}}{12167}+\frac{1890\pi\ln^2\left(\frac{x}{3}\right)}{12167}\right)\bigg]\frac{1}{x^3}+O\left(\frac{1}{x^4}\right)$}
\end{split}\end{equation}
this solution holds for $k\ne0$ and the Gaussian regime is cut in the large $x$ limit. This expansion, with $x_{\mathrm{BC}}\to\infty$, is used in eq. (\ref{UVefateff}) and eq. (\ref{serieUVgeff}).

\subsection{The perturbations around the gaussian fixed-point}
The perturbations of the form factors around the gaussian fixed-point for $g_k$ are given by 
\begin{equation}
\label{pertGFPF}
F_k\left(x;x_0,\frac{k}{m_p}\ll1\right)=F_k^{\mathrm{GFP}}\left(x;x_0\right)=F_\ast^{\mathrm{GFP}}\left(x;x_0\right)+\sum_{j=1}^{+\infty}{F_j^{\mathrm{GFP}}\left(x;x_0\right)\left(\frac{k}{m_p}\right)^{2j}}
\end{equation}
The general functional form is given by
\begin{equation}
F_j^{\mathrm{GFP}}=a_jx^j\ln{\left(\frac{x}{x_0}\right)}+A_j\left(x;x_0\right)+B_j\left(x\right)\tanh^{-1}\left(\frac{\sqrt x}{\sqrt{x+12}}\right)+C_j\left(x_0\right)\tanh^{-1}\left(\frac{\sqrt{x_0}}{\sqrt{x_0+12}}\right)
\end{equation}
The gaussian fixed-point is given by
\begin{equation}\begin{split}
&F_\ast^{\mathrm{GFP}}\left(x;x_0\right)=\frac{\left(x-x_0\right)\left(\left(x_0\left(97x_0+480\right)+432\right)x^2+48\left(x_0+12\right)\left(10x_0+9\right)x+432x_0\left(x_0+12\right)\right)}{320\pi^2x^2\left(x+12\right)x_0^2\left(x_0+12\right)}\\
&-\frac{7\log{\left(\frac{x}{x_0}\right)}}{320\pi^2}+\frac{7x\left(x\left(x\left(x+18\right)+54\right)+432\right)+2592}{160\pi^2x^\frac{5}{2}\left(x+12\right)^\frac{3}{2}}\tanh^{-1}\left(\frac{\sqrt x}{\sqrt{x+12}}\right)\\
&+\frac{-7x_0\left(x_0\left(x_0\left(x_0+18\right)+54\right)+432\right)-2592}{160\pi^2x_0^{\frac{5}{2}}\left(x_0+12\right)^\frac{3}{2}}\tanh^{-1}\left(\frac{\sqrt{x_0}}{\sqrt{x_0+12}}\right)
\end{split}\end{equation}
The perturbation with $j=1$ is given by
\begin{equation}\begin{split}
&F_{j=1}^{\mathrm{GFP}}\left(x;x_0\right)=\frac{23\left(-\frac{2\left(187x+900\right)}{x^2}-\frac{157x}{x_0}-\frac{105x}{x_0+12}+\frac{374x}{x_0^2}+\frac{1800x}{x_0^3}-\frac{1260}{x+12}+262\right)}{17920\pi^3}+\frac{3013x\log{\left(\frac{x}{x_0}\right)}}{53760\pi^3}\\
&+\frac{8942400-23x\left(131x\left(x+6\right)\left(x\left(x+12\right)-18\right)-134784\right)}{26880\pi^3x^\frac{5}{2}\left(x+12\right)^\frac{3}{2}}\tanh^{-1}\left(\frac{\sqrt x}{\sqrt{x+12}}\right)\\
&+\frac{23x\left(x_0\left(131x_0\left(x_0+6\right)\left(x_0\left(x_0+12\right)-18\right)-134784\right)-388800\right)}{26880\pi^3x_0^{\frac{7}{2}}\left(x_0+12\right)^\frac{3}{2}}\tanh^{-1}\left(\frac{\sqrt{x_0}}{\sqrt{x_0+12}}\right)
\end{split}\end{equation}
The perturbation with $j=2$ is given by
\begin{equation}\begin{split}
&F_{j=2}^{\mathrm{GFP}}\left(x;x_0\right)=\\
&=\frac{529}{193536\pi^4\left(x+12\right)x_0^4\left(x_0+12\right)x^2}\Bigg(-\left(\left(x_0\left(x_0\left(55x_0\left(2x_0+15\right)-1821\right)+12384\right)+45360\right)x^5\right)+\\
&+2\left(x_0\left(x_0\left(55x_0\left(x_0^2-90\right)+10926\right)-74304\right)-272160\right)x^4+825x_0^4\left(x_0+12\right)x^3-\\
&-1821x_0^4\left(x_0+12\right)x^2+12384x_0^4\left(x_0+12\right)x+45360x_0^4\left(x_0+12\right)\Bigg)+\frac{29095x^2\log{\left(\frac{x}{x_0}\right)}}{580608\pi^4}\\
&-\frac{529\left(x\left(55x\left(x\left(x+6\right)\left(x\left(x+12\right)-18\right)+486\right)+268272\right)+816480\right)}{290304\pi^4\left(x+12\right)^\frac{3}{2}x^\frac{5}{2}}\tanh^{-1}\left(\frac{\sqrt x}{\sqrt{x+12}}\right)\\
&+\frac{529\left(x_0\left(55x_0\left(x_0\left(x_0+6\right)\left(x_0\left(x_0+12\right)-18\right)+486\right)+268272\right)+816480\right)x^2}{290304\pi^4x_0^{\frac{9}{2}}\left(x_0+12\right)^\frac{3}{2}}\tanh^{-1}\left(\frac{\sqrt{x_0}}{\sqrt{x_0+12}}\right)
\end{split}\end{equation}
The parameter $x_0$ sets the initial scale for the Cauchy problem associated with the perturbative equations. Its value is determined numerically by matching the perturbative expansion with the asymptotic regimes of the full numerical solution in $k$ given by eq.(\ref{ASsoladim}). These expansions are used in eq. (\ref{etaeffGFP}) and eq. (\ref{geffGFP}) with $x_0=1$ for the illustration.

Rewriting eq.(\ref {ASsoladim}) as
\begin{equation}
F_k(x;\Lambda) =\int_{\frac{k^2x}{\Lambda^2}}^{x_{BC}}{\frac{H\left(\eta_\ast=-2,v\right)}{2v}dv}-\int_{\frac{k^2x}{\Lambda^2}}^{x}{\frac{H\left(\eta\left(k\sqrt{\frac{x}{v}}\right),v\right)}{2v}dv}
\end{equation}
and expanding around $k\to0$, formally the second integral reproduces the same structure as the fixed-point expansion discussed above, with $x_0\equiv k^2x/\Lambda^2$. The limit $\Lambda\to\infty$ is then equivalent to consider $x_0\to0$. The expansion of $F_k(x;\Lambda)$ reduces to
\begin{equation}
\lim_{\Lambda\rightarrow\infty}{F_k\left(x,\frac{k}{m_p}\ll1;\Lambda\right)}=F_k^{GFP}\left(x\right)+\sum_{j=0}^{+\infty}{\left[a_j(x_{BC})+b_j(x_{BC})\ln{\left(\frac{23}{12\pi}\frac{k^2x}{m_p^2}\right)}\right]\left(\frac{k^2}{m_p^2}\right)^j}
\end{equation}
The second term corresponds to $F_B(k^2/m_p^2)$ in eq.~(\ref{serieGFPNGFP}). $F_B$ generates a residual logarithmic contribution, preventing the integrated solution from reducing to the stationary Gaussian scaling solution in the infrared limit. In the limit $x\to\infty$, along with $x_{\mathrm{BC}}\to\infty$, this expansion reproduces eq.(\ref{formASIR}).

\subsection{Perturbations around the non-trivial fixed point}
The perturbations for the form factor around the non-trivial fixed-point of $g_k$ take the form:
\begin{equation}
\label{pertNGFPF}
F_k\left(x;x_0,\frac{k}{m_p}\gg1\right)=F_\ast^{\mathrm{NGFP}}\left(x;x_0\right)+\sum_{n=1}^{+\infty}{F_j^{\mathrm{NGFP}}\left(x;x_0\right)\left(\frac{m_p}{k}\right)^{2j}}
\end{equation}
The general functional form is given by
\begin{equation}
F_j^{\mathrm{NGFP}}=a_j\delta_{j,0}\ln{\left(\frac{x}{x_0}\right)}+A_j\left(x;x_0\right)+B_j\left(x\right)\tanh^{-1}\left(\frac{\sqrt x}{\sqrt{x+12}}\right)+C_j\left(x_0\right)\tanh^{-1}\left(\frac{\sqrt{x_0}}{\sqrt{x_0+12}}\right)
\end{equation}
The non-gaussian fixed-point is given by
\begin{equation}\begin{split}
&F_\ast^{\mathrm{NGFP}}\left(x;x_0\right)=\frac{\left(x-x_0\right)\left(\left(x_0\left(29x_0+150\right)+324\right)x^2+6\left(x_0+12\right)\left(25x_0+54\right)x+324x_0\left(x_0+12\right)\right)}{40\pi^2x^2\left(x+12\right)x_0^2\left(x_0+12\right)}+\\
&-\frac{7\log{\left(\frac{x}{x_0}\right)}}{120\pi^2}+\frac{7x\left(x\left(x\left(x+18\right)+54\right)+432\right)+5832}{60\pi^2x^\frac{5}{2}\left(x+12\right)^\frac{3}{2}}\tanh^{-1}\left(\frac{\sqrt x}{\sqrt{x+12}}\right)\\
&+\frac{-7x_0\left(x_0\left(x_0\left(x_0+18\right)+54\right)+432\right)-5832}{60\pi^2x_0^{\frac{5}{2}}\left(x_0+12\right)^\frac{3}{2}}\tanh^{-1}\left(\frac{\sqrt{x_0}}{\sqrt{x_0+12}}\right)\ 
\end{split}\end{equation}
The perturbation with $j=1$ is given by
\begin{equation}\begin{split}
&F_{j=1}^{\mathrm{NGFP}}\left(x;x_0\right)=\frac{3\left(x-x_0\right)\left(x\left(13x_0-60\right)-60\left(x_0+12\right)\right)}{92\pi x^2\left(x+12\right)x_0\left(x_0+12\right)}+\\
&+\frac{-x\left(13x\left(x+18\right)+2592\right)-4320}{184\pi x^\frac{5}{2}\left(x+12\right)^\frac{3}{2}}\tanh^{-1}\left(\frac{\sqrt x}{\sqrt{x+12}}\right)\\
&+\frac{x_0\left(13x_0\left(x_0+18\right)+2592\right)+4320}{184\pi x x_0^{\frac{3}{2}}\left(x_0+12\right)^\frac{3}{2}}\tanh^{-1}\left(\frac{\sqrt{x_0}}{\sqrt{x_0+12}}\right)
\end{split}\end{equation}
The perturbation with $j=2$ is given by
\begin{equation}\begin{split}
&F_{j=2}^{\mathrm{NGFP}}\left(x;x_0\right)=+\frac{\left(x-x_0\right)\left(156x_0+13x\left(x_0+12\right)+4464\right)}{2116x^2\left(x+12\right)\left(x_0+12\right)}-\\
&-\frac{9\left(x\left(13x+24\right)-720\right)}{529x^\frac{5}{2}\left(x+12\right)^\frac{3}{2}}\tanh^{-1}\left(\frac{\sqrt x}{\sqrt{x+12}}\right)+\frac{9\left(x_0\left(13x_0+24\right)-720\right)}{529x^2\sqrt{x_0}\left(x_0+12\right)^\frac{3}{2}}\tanh^{-1}\left(\frac{\sqrt{x_0}}{\sqrt{x_0+12}}\right)
\end{split}\end{equation}
For the fixed point, matching with eq.~(\ref{ASsoladim}) fixes $x_0 = x_{\mathrm{BC}}$. In the perturbative expansion, the matching scale is instead identified with $x_0 = q^2/\Lambda^2$, implying $x_0 \to 0$ and thus $x_0^{(j)} = 0$ for $j \ge 1$ as $\Lambda\to\infty$. These limits, together with $x_{\mathrm{BC}}\to\infty$, are used in eqs.~(\ref{etaeffNGFP}) and (\ref{geffNGFP}).

For all $j\ge1$, the limit $x_0\to0$ is smooth, and $F_j^{\mathrm{NGFP}}$ reduces to a function of $x$ only, eq.~(\ref{serieGFPNGFP}). The subsequent limit $x\to\infty$ reproduces eq.~(\ref{formASUV}).

\end{appendices}

\bibliography{refformfact}
\end{document}